\documentclass[12pt]{article}

\usepackage{amsmath,amsthm,amssymb,amsfonts}
\usepackage{graphics,graphicx,color}

\graphicspath{{/EPSF/}{Figures/}}
\DeclareGraphicsExtensions{.eps}
\DeclareGraphicsExtensions{.pdf,.jpg}

\makeatletter

\@addtoreset{equation}{section}
\makeatother

\newtheorem{theorem}{Theorem}[section]
\newtheorem{lemma}[theorem]{Lemma}
\newtheorem{proposition}[theorem]{Proposition}

\newtheorem{remark}[theorem]{Remark}

\newcommand{\dint}{\displaystyle\int}

\def\ep{\varepsilon}
\def\v{\varphi}

\def\R{\mathbb R}
\def\S{\mathbb S}
\def\N{\mathbb N}
\def\pa{\partial}

\def\A{{\mathcal A}}
\def\v{\mathrm v}
\def\V{\mathrm V}
\def\Tu{\mathbf{T_1}}
\def\T{\mathbf{T}}

\begin{document}

\title{Stability estimates in stationary inverse transport}

\author{Guillaume Bal and Alexandre Jollivet 
\thanks{Department of Applied Physics and 
        Applied Mathematics, Columbia University, 
        New York,  NY 10027, USA; gb2030@columbia.edu and aj2315@columbia.edu}
}

\maketitle

\begin{abstract}
  We study the stability of the reconstruction of the scattering and
  absorption coefficients in a stationary linear transport equation
  from knowledge of the full albedo operator in dimension $n\geq3$.
  The albedo operator is defined as the mapping from the incoming
  boundary conditions to the outgoing transport solution at the
  boundary of a compact and convex domain. The uniqueness of the
  reconstruction was proved in \cite{choulli-stefanov-IP,CS-OSAKA-99}
  and partial stability estimates were obtained in \cite{W-AIHP-99}
  for spatially independent scattering coefficients. We generalize
  these results and prove an $L^1$-stability estimate for spatially
  dependent scattering coefficients.
\end{abstract}

\section{Introduction}

Let the spatial domain $X\subset \R^n$, $n\ge 2$, be a convex, open
bounded subset with $C^1$ boundary $\pa X$, and let the velocity
domain $V$ be $\S^{n-1}$ or an open subset of $\R^n$ which satisfies
$\inf_{v\in V}|v|>0$.  Let $\Gamma_{\pm}=\{(x,v)\in \pa X\times V; \pm
n(x)v>0\}$ where $n(x)$ denotes the outward normal vector to $\pa X$
at $x\in \pa X$.  The set $\Gamma_-$ is the set of incoming boundary
condition while $\Gamma_+$ is the set where we measure the outgoing
solution to the following stationary linear Boltzmann transport
equation in $X\times V$:
\begin{eqnarray}
&&\!\!\!\!\!\!\!\!\!\!\!\!\!\!v\nabla_xf(x,v)+\sigma(x,v)f(x,v)-\int_Vk(x,v',v)f(x,v')dv'=0
\quad\textrm{in } X\times V,\label{eq:tr}\\
&&\!\!\!\!\!\!\!\!\!\!\!\!\!\!f_{|\Gamma_-}=f_-.\nonumber
\end{eqnarray}
Here, $f(x,v)$ models the density of particles at position $x\in X$ with
velocity $v\in V$.

The albedo operator $\mathcal A$ is then defined by
\begin{equation}
  \label{eq:albedo}
  \mathcal A: f_- \mapsto f_{|\Gamma_+},
\end{equation}
where $f(x,v)$ is the solution to \eqref{eq:tr}. The inverse transport
problem consists of reconstructing the absorption coefficient
$\sigma(x,v)$ and the scattering coefficient $k(x,v',v)$ from
knowledge of $\mathcal A$. Stability estimates aim at controlling the
variations in the reconstructed coefficients $\sigma(x,v)$ and
$k(x,v',v)$ from variations in $\mathcal A$ in suitable metrics.

The forward transport equation has been analyzed in e.g.
\cite{dlen6,mokhtar,RS-79-III}. The inverse transport problem has been
addressed in e.g.
\cite{choulli-stefanov-IP,CS-OSAKA-99,Rom-JIIPP-97,S-IO-03} with
stability estimates obtained in \cite{Rom-JIIPP-97,W-AIHP-99}.  For
the two-dimensional case, in which proofs of uniqueness of the
scattering coefficient are available only when it is sufficiently
small or independent of the spatial variable, we refer the reader to
e.g. \cite{IP2-00,SU-MAA-03,tamasan-IP02}.

To obtain our stability estimates, we follow a methodology based on
the decomposition of the albedo operator into singular components
\cite{choulli-stefanov-IP,CS-OSAKA-99} and the use of appropriate
functions on $\Gamma_\pm$ with decreasing support \cite{W-AIHP-99}. In
dimensions $n\geq3$ the contribution due to single scattering is more
singular than the contribution due to higher orders of scattering. As
a consequence, the single scattering in a direction $v'$ generated by
a delta function $f_-=\delta_{x_0}(x)\delta(v-v_0)$ is a
one-dimensional curve on $\partial X$. In order to obtain general
stability estimates for the scattering coefficient, one way to proceed
is to construct test functions whose support converges to that
specific curve. It turns out that it is simpler to work in a geometry
in which this curve becomes a straight line.

We now briefly introduce that geometry and refer the reader to section
\ref{sec:albedo} below for a formal presentation. Let $R$ be a
positive real constant such that $X$ is included in the ball $B(R)$ of
radius $R$ centered at $x=0$.  On $(B(R)\backslash X)\times V$, the
absorption and scattering coefficients vanish and we may solve the
equation $v\nabla_x f=0$.  This allows us to map back the incoming
conditions $f_-$ on $\Gamma_-$ as incoming conditions, which we shall
still denote by $f_-$, on $F_-$ and map forward the outgoing solution
$f_{|\Gamma_+}$ to an outgoing solution $f_+$ on $F_+$, where we have
defined
\begin{equation}
  \label{eq:Fpm}
  F_\pm := \{ (x\pm R\hat v,v)\in \R^n\times V \mbox{ for } 
  (x,v)\in\R^n\times V\mbox{ s.t. } vx=0,\,|x|<R\}.
\end{equation}
In other words, $F_\pm$ is the union for each $v\in V$ of
the spatial points on a disc of radius $R$ in a plane 
orthogonal to $v$ and tangent to the sphere of radius $R$.

\begin{figure}[h]
  \centering
  \includegraphics[width=8cm]{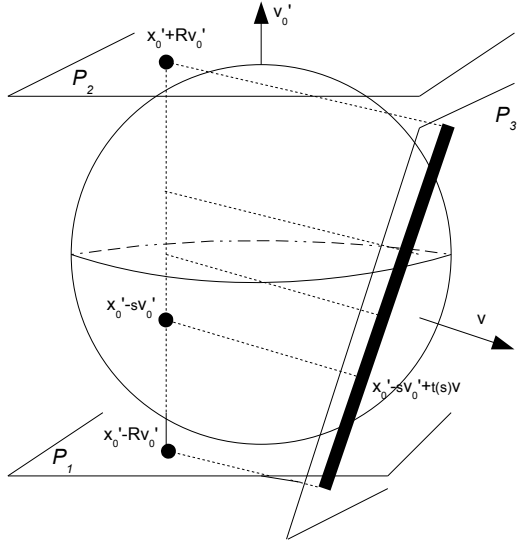}
  \caption{Geometry of the ballistic and single scattering components 
    in dimension $n=3$.  The source term is non-zero in the vicinity
    (in $F_-$) of $x_0'-Rv_0'$ in $P_1=-Rv'_0+\Pi_{v'_0}(R)$. The
    ballistic part is non-zero in the vicinity (in $F_+$) of
    $x_0'+Rv_0'$ in $P_2=Rv'_0+\Pi_{v'_0}(R)$. The thick ``line''
    represents the support of the single scattering contribution in
    the vicinity (in $F_+$) of the segment $\{x_0'-sv_0'+t(s)v;s\in
    (-R,R)\}\subset P_3=Rv+\Pi_v(R)$. See text for the notation.}
  \label{fig:1}
\end{figure}

The incoming boundary condition is thus now defined on $F_-$ while
measurements occur on $F_+$ and we may define the albedo operator
still called $\mathcal A$ as an operator mapping $f_-$ defined on
$F_-$ to the outgoing solution $f_+$ on $F_+$. We may now verify that
the single scattering in a direction $v'$ generated by a delta
function $f_-=\delta(x-x_0)\delta(v-v_0)$ for $(x_0,v_0)\in F_-$ is a
one-dimensional segment in $F_+$; see Fig. \ref{fig:1}. Note also that
the geometry we consider here may be more practical than the geometry
based on $\Gamma_\pm$. Indeed, we assume that the incoming conditions
are generated on a plane for each direction of incidence, and, more
importantly, that our measurements are acquired on a plane for each
outgoing direction. This is how the collimators used in Computerized
Tomography \cite{NW-SIAM01} are currently set up.

Under appropriate assumptions on the coefficients, we aim to show that
$\mathcal A$ is a well posed operator from $L^1(F_-)$ to $L^1(F_+)$.
We shall then obtain a stability estimate for the reconstruction
$\sigma(x)$ (or $\sigma(x,|v|)$) and $k(x,v,v')$ with respect to the
norm ${\mathcal L}(L^1(F_-),L^1(F_+))$ of $\mathcal A$.

The rest of the paper is structured as follows.  Because our geometry
is not standard, we present a detailed analysis of the linear
transport equation and of the singular decomposition of the albedo
operator in section \ref{sec:albedo}. Most of the material in that
section is similar to that in \cite{CS-OSAKA-99}. One of the main
physical constraints in the existence of solutions to \eqref{eq:tr} is
that the system be ``subcritical'', in the sense that the
``production'' of particles by the scattering term involving the
scattering coefficient $k(x,v,v')$ has to be compensated by the
absorption of particles and the leakage of particles at the domain's
boundary. Although this may be seen implicitly in \cite{CS-OSAKA-99},
we state explicitly that the decomposition of the albedo operator used
in the stability estimates holds as soon as the forward transport
problem is well-posed in a reasonable way.

The stability results are stated in section \ref{sec:stab}. Under
additional continuity assumptions on the absorption and scattering
coefficients, we obtain that (i) the exponential of line integrals of
the absorption coefficient and (ii) the scattering coefficient
multiplied by the exponential of the integral of the absorption
coefficient on a broken line are both stably determined by $\mathcal
A$ in ${\mathcal L}(L^1(F_-),L^1(F_+))$; see Theorem \ref{thm:stab1}.
Under additional regularity hypotheses on the absorption coefficient,
we obtain a stability result for the absorption coefficient in some
Sobolev space $H^s$ and for the scattering coefficient in the $L^1$
norm. The stability results in the geometry of \eqref{eq:tr} are
presented in section \ref{sec:stab2}.  The proof of the stability
results and the construction of the appropriate test functions are
presented in section \ref{sec:proof3}.  Several proofs on the
decomposition of the albedo operator and the uniqueness of the
transport equation have been postponed to sections \ref{sec:proof2}
and \ref{sec:proof1}, respectively.

\section{Transport equation and albedo operator}
\label{sec:albedo}

We now state our main results on the stationary linear transport equation
and the corresponding albedo operator. 
%We decompose the albedo operator
%following \cite{CS-OSAKA-99}.

%\subsection{Stationary linear transport equation}

Let $R$ be a positive real constant and let $n\in \N$, $n\ge 2$. Let
$V$ be $\S^{n-1}$ or an open subset of $\R^n$ which satisfies
$\v_0:=\inf_{v\in V}|v|>0$.  For $v\in V$, we define
\begin{math}
  \hat v:={v\over |v|}.
\end{math}
%\begin{equation}
%\hat v:={v\over |v|}.\label{0.0}
%\end{equation}
Then, we consider the open subset $O$ of $\R^n\times V$ defined by
\begin{equation}
O:=\{(x,v)\in \R^n\times V\ |\ |x\hat v|<R,\ |x-(x\hat v)\hat v|<R\},
  \label{0.1}
\end{equation}
and let $F$ be the set
\begin{equation}
F:=\{(x,v)\in \R^n\times V\ |\ x\in \Pi_v(R)\},\label{0.2}
\end{equation}
where 
\begin{equation}
\Pi_v(R):=\{x\in \R^n\ | \ xv=0,\ |x|<R\},\label{0.2a} 
\end{equation}
for all $v\in V$. For all $v\in V$ we also consider
\begin{equation} 
R\hat v+\Pi_v(R):=\{R\hat v+x\ | \ x\in \Pi_v(R)\}.\label{0.2b} 
\end{equation}

When $V=\S^{n-1}$, then $F$ is an open subset of
$T\S^{n-1}:=\{(x,v)\in \R^n\times \S^{n-1}\ |\ vx=0\}$, the tangent
space to the unit sphere.  When $V$ is an open subset of $\R^n$ (which
satisfies $\v_0=\inf_{v\in V}|v|>0$) then $F$ is an open subset of the
$2n-1$ dimensional manifold $\{(x,v)\in \R^n\times V\ |\ vx=0\}$.  We
also define $F_\pm$ by
\begin{equation}
F_\pm:=\{(x\pm R\hat v,v)\in \R^n\times V| (x,v)\in F\},\label{0.3}
\end{equation}
and recall that $F_-$ is the set of incoming conditions for the transport
equation while $F_+$ is the set in which measurements are performed.

We consider the space $L^1(O)$ with the usual norm
\begin{equation}
\|f\|_O:=\int_O|f(x,v)|dx dv,\textrm{ for } f\in L^1(O).
\end{equation}
We also consider the space $L^1(F)$ defined as the completed Banach
space of the vector space of compactly supported continuous functions
on $F$ for the norm
\begin{equation}
\|f\|_F:=\int_V\int_{\Pi_v(R)}|v||f(x,v)|dxdv,\ f\in L^1(F),
\end{equation}
and similarly the spaces $L^1(F_\pm)$ defined as the completed Banach
space of the vector space of compactly supported continuous functions
on $F_\pm$ for the norm
\begin{equation}
\|f_\pm\|_{F_\pm}:=\int_V\int_{\Pi_v(R)}|v||f_\pm(x\pm R\hat v,v)|dxdv,
\,f_\pm\in L^1(F_\pm).
\end{equation}

We assume that:
\begin{equation}
  \label{eq:hyp1}
  \begin{array}{l}
0\le \sigma\in L^{\infty}(\R^n\times V),\\
k(x,v',v)\textrm{ is a measurable function on }\R^n\times V\times V, \\
\sigma(x,v)=k(x,v',v)=0 \textrm{ for } (x,v',v)\in \R^n\times V\times V,\ |x|>R,\\
0\le k(x,v',.)\in L^1(V)\textrm{ for a.e. }(x,v')\in \R^n\times V\textrm{ and }\\
\sigma_p(x,v')=\int_Vk(x,v',v)dv\textrm{ belongs to } L^{\infty}(\R^n\times V).
  \end{array}
\end{equation}
Under these conditions, we consider the stationary linear Boltzmann
transport equation
\begin{eqnarray}
&&v\nabla_xf(x,v)+\sigma(x,v)f(x,v)-\int_Vk(x,v',v)f(x,v')dv'=0
  \,\,\textrm{ in }O,\label{0.8}\\
&&f_{|F_-}=f_-.\nonumber
\end{eqnarray}
Throughout the paper, for $m\in \N$ and for any subset $U$ of $\R^m$
we denote by $\chi_U$ the characteristic function defined by
$\chi_U(x)=0$ if $x\not\in U$ and $\chi_U(x)=1$ if $x\in U$.

%\subsection{Trace results and  change of variables}
We now analyze the well-posedness of \eqref{0.8}.  The following
change of variables is useful.
\begin{lemma}
  \label{lem:chvar}
  For $f\in L^1(O)$, we have:
\begin{equation}
\int_Of(x,v)dx dv=\int_V\int_{\Pi_v(R)}\int_{-R}^Rf(y\pm t\hat v,v)dt dydv.
\label{2.1}
\end{equation} 
\end{lemma}
\begin{proof}%[Proof of Lemma 2.1]
  First using \eqref{0.1} we have
  $$
  \int_Of(x,v)dx dv=\int_V\int_{\R^n}\chi_O(x,v)f(x,v)dx dv.
  $$
  Then for a.e. $v\in V$ using the change of variables $\{x\in\R^n\ 
  | \ |\hat v.x|<R,\ |x-(\hat vx)\hat v|<R\}\to \Pi_v(R)\times (-R,R)$, $x\mapsto (x-(\hat vx)\hat v, \pm \hat vx)$,
  we obtain \eqref{2.1}.
\end{proof}

We introduce the following notation:
\begin{eqnarray}
&&T_0f=-v\nabla_x f\ \textrm{in the distributional sense},\ \ A_1f=-\sigma f,\label{eq:opTA}\\
&& A_2f=\int_Vk(x,v',v)f(x,v')dv',\ T_1=T_0+A_1,\ T=T_0+A_1+A_2=T_1+A_2, \nonumber
\end{eqnarray}
and the Banach spaces
\begin{displaymath}
  \begin{array}{rclrcl}
  {\mathcal W}&:=&\{f\in L^1(O,|v|dx dv); T_0f\in L^1(O)\},\,
   &\|f\|_{{\mathcal W}}&=&\|T_0f\|_O+\||v|f\|_O,\\
 \tilde{\mathcal W}&:=&\{f\in L^1(O); T_0f\in L^1(O)\},\,&
\|f\|_{\tilde{\mathcal W}}&=&\|T_0f\|_O+\|f\|_O.
  \end{array}
\end{displaymath}
We consider the space $L(F_\pm)$ defined as the completed Banach space
of the vector space of compactly supported continuous functions on
$F_\pm$ for the norm
\begin{equation}
\|f_\pm\|_{L(F_\pm)}:=\int_V\int_{\Pi_v(R)}|f_\pm(x'\pm R\hat v,v)|dx'dv,\textrm{ for }f_\pm\in L(F_\pm).
\end{equation}
Note that ${\mathcal W}\subseteq \tilde {\mathcal W}$ and
$L^1(F_{\pm})\subseteq L(F_{\pm})$. The spaces $\tilde{\mathcal W}$ and
$L(F_\pm)$ are used only to define the unbounded operators $\T$ and
$\Tu$ below. We obtain the following trace result.
\begin{lemma}
  \label{lem:trace}
  We have
  \begin{equation}
 \|f_{|F_\pm}\|_{F_\pm}\le C\|f\|_{\mathcal W},\label{2.2}
  \end{equation}
  for $f\in {\mathcal W}$, where $C=\max((2R)^{-1}, 1)$ and
  \begin{equation}
\|f_{|F_\pm}\|_{L(F_\pm)}\le C'\|f\|_{\tilde{\mathcal W}},\label{2.2'}
  \end{equation}
  for $f\in \tilde{\mathcal W}$, where $C'=\max((2R)^{-1},\v_0^{-1})$.
\end{lemma}
\begin{proof}%[Proof of Lemma 2.2]
  Let $f$ be a $C^1$ function in $\R^n\times V$ with compact support.
  Then from \eqref{2.1}, it follows that
\begin{equation}
\|f\|_{{\mathcal W}}=\int_V\int_{\Pi_v(R)}\int_{-R}^R(|v||{d\over dt}f(x'\pm t\hat v,v)|+|v||f(x'\pm t\hat v,v)|)dt dx'dv.\label{2.3'}
\end{equation}
Let $v\in V$ and $x'\in \R^n$ such that $vx=0$. Note that $f(x'\mp
R\hat v,v)=f(x'\pm t\hat v,v)-\int_{-R}^t{d\over ds}f(x'\pm s\hat
v,v)ds$ for all $t\in (-R,R)$. Hence $|f(x'\mp R\hat v,v)|\le |f(x'\pm
t\hat v,v)|+\int_{-R}^R|{d\over ds}f(x'\pm s\hat v,v)|ds$. Upon
integrating the latter equality, we obtain
\begin{equation}
|f(x'\mp R\hat v,v)|\le {1\over 2R}\int_{-R}^R|f(x'\pm t\hat v,v)|dt+\int_{-R}^R|{d\over ds}f(x'\pm s\hat v,v)|ds.\label{2.4'}
\end{equation}
Combining \eqref{2.3'} et \eqref{2.4'}, we obtain \eqref{2.2}.  The
proof of \eqref{2.2'} is similar.
\end{proof}
For a continuous function $f_-$ on $F_-$, we define the following
extension of $f_-$ in $O$:
\begin{equation}
Jf_-(x,v)=e^{-|v|^{-1}\int_0^{R+x\hat v}\sigma (x-s\hat v,v)ds }f_-(x-(x\hat v+R)\hat v,v),\ (x,v)\in O.\label{2.3}
\end{equation}
\begin{lemma}
  \label{lem:estJ}
  For $f_-\in L^1(F_-)$ with $C=2R(1+\v_0^{-1}\|\sigma\|_{\infty})$, we have:
  %The following estimate is valid :
  \begin{equation}
  \|Jf_-\|_{\mathcal W}\le C\|f_-\|_{F_-}.\label{2.4}
  \end{equation}
  %for $f_-\in L^1(F_-)$, where $C=2R(1+\v_0^{-1}\|\sigma\|_{\infty})$.
\end{lemma}
\begin{proof}%[Proof of Lemma 2.3]
  Let $f_-$ be a compactly supported continous function on $F_-$.
  From \eqref{2.1} and \eqref{2.3} it follows that
\begin{equation}
\||v|Jf_-\|_O\!=\!\!\int_V\int_{\Pi_v(R)}\!\!\!\!\!\!\!\!\!\!\!\!|v||f_-(x'-R\hat v,v)|\int_{-R}^R\!\!\!\!\!\!\!e^{-|v|^{-1}\int_{-R}^t\sigma(x'+s\hat v,v)ds}dtdx'dv\le 2R\|f_-\|_{F_-}\label{2.5}.
\end{equation}
One can check that $Jf_-$ satisfies 
\begin{math}
  T_0Jf_-=-A_1Jf_-
\end{math}
in the distributional sense.  Therefore using also \eqref{2.5} we
obtain
$\|T_0Jf_-\|_O+\||v|Jf_-\|_O\le (1+\|A_1|v|^{-1}\|)$ $\times\||v|Jf_-\|_O=2R(1+\v_0^{-1}\|\sigma\|_{\infty})\|f_-\|_{F_-}$, which proves the lemma. %Lemma 2.3.
\end{proof}

\subsection{Existence theory for the albedo operator}

We consider the following unbounded operators:
\begin{equation}
\Tu f=T_1f,\ \T f=Tf,\ D(\Tu)=D(\T)=\{f\in \tilde{\mathcal W}\ ; \ f_{|F_-}= 0\}. 
\end{equation}
The operator $\Tu:D(\Tu)\to L^1(O)$ is close, one-to-one, onto, and
its inverse $\Tu^{-1}$ is given for all $f\in L^1(O)$ by
\begin{equation}
\Tu^{-1}f(x,v)=-|v|^{-1}\int_0^{R+x\hat v}e^{-|v|^{-1}\int_0^t\sigma(x-s \hat v,v)ds}f(x-t\hat v,v)dt,\ (x,v)\in O.\label{tu}
\end{equation}
%for all $f\in L^1(O)$. 
\begin{lemma}
  \label{lem:exist}
  The following statements hold:
\begin{itemize}
\item[i.]The bounded operator $|v|\Tu^{-1}$ in $L^1(O)$ has norm less or equal to $2R$ and the bounded operator $A_2|v|^{-1}$ in $L^1(O)$ has norm less than $\||v|^{-1}\sigma_p(x,v)\|_{L^\infty(O)}$.
\item[ii.]Under the hypothesis
\begin{equation}
\sigma-\sigma_p\ge 0,\label{c2}
\end{equation}
the bounded operator $A_2\Tu^{-1}$ in $L^1(O)$ has norm less than $1-e^{-2R\v_0^{-1}\|\sigma_p\|_\infty}$.
\item[iii.]Assume either condition \eqref{c2} or
\begin{equation}
2R\||v|^{-1}\sigma_p(x,v)\|_{L^\infty(O)}<1.\label{c1}
\end{equation}
Then $I+A_2\Tu^{-1}$ is invertible in $L^1(O)$.
\end{itemize}
\end{lemma}
Lemma \ref{lem:exist} is proved in section 7.  We denote by $K$ the
bounded operator in $L^1(O, |v|dx dv)$ defined by $K=\Tu^{-1}A_2$:
$$
Kf(x,v)=-|v|^{-1}\int_0^{R+x\hat v}e^{-|v|^{-1}\int_0^t\sigma(x-s\hat v,v)ds}(A_2f)(x-t\hat v,v)dt,\ (x,v)\in X\times V,
$$
for all $f\in L^1(O, |v| dx dv)$. The operator $K$ also defines a
bounded operator in $L^1(O)$.  This allows us to recast the
stationary linear Boltzmann transport equation as  the following
integral equation:
\begin{equation}
(I+K)f=Jf_-.\label{3.1}
\end{equation}
The existence theory for the above integral equation is addressed in
the following result.
\begin{proposition}
  \label{prop:fwd}
  The following statements hold:
  \begin{itemize}    
  \item[i.]The conditions \eqref{c4} and \eqref{c5} below are
    equivalent.
  \begin{eqnarray}
&&\textrm{The bounded operator }I+K \textrm{ in } L^1(O) \textrm{ admits a bounded}
\nonumber\\&&
\textrm{inverse in } L^1(O).\label{c4}\\ 
&&\textrm{The bounded operator }I+A_2\Tu^{-1} \textrm{ in } L^1(O) \textrm{ admits a bounded}\nonumber\\
&&\textrm{inverse in } L^1(O).\label{c5}
  \end{eqnarray}
  
\item[ii.]Assume either \eqref{c2} or \eqref{c1}. Then condition
  \eqref{c4} is satisfied.
  
\item[iii.]If \eqref{c4} is satisfied then
  \begin{eqnarray}
  &&\textrm{the bounded operator }I+K \textrm{ in } L^1(O,|v|dxdv)\nonumber\\
 && \textrm{admits a bounded inverse in } L^1(O,|v|dxdv).\label{c3}
  \end{eqnarray}
  \end{itemize}
\end{proposition}
Proposition \ref{prop:fwd} is proved in Section 7.  The following
proposition deals with the existence of the albedo operator.
\begin{proposition}
  \label{prop:albedo}
  Assume \eqref{c3}.  Then
  \begin{itemize}
  \item[i.]the integral equation \eqref{3.1} is uniquely solvable for
    all $f_-\in L^1(F_-)$ and $f\in {\mathcal W}$ ;
    
  \item[ii.]the albedo operator ${\mathcal A}: f_-\mapsto f_+=f_{|F_+}$ is
    a bounded operator ${\mathcal A}: L^1(F_-)\to L^1(F_+)$.
\end{itemize}
\end{proposition}
Proposition \ref{prop:albedo} is proved in Section 7.  
%\vskip 2mm {\bf
%  Remark 2.1.}  Under condition \eqref{c4} one can also define the
%albedo operator from $L(F_-)$ to $L(F_+)$.

\subsection{Singular decomposition of the albedo operator}
We assume that condition \eqref{c3} is satisfied.  Let us consider the
operator 

\noindent ${\mathcal R}:L^1(O, |v|dx dv)\to L^1(F_+),$ defined by
\begin{equation}
\psi\mapsto {\mathcal R}\psi := \left(K^2\psi\right)_{|F_+},\label{S0}
\end{equation} 
for $\psi\in L^1(O, |v|dx dv)$.  Using the equality (in the
distributional sense) $T_0Kf=-A_1f-A_2f$ for $f\in L^1(O, |v|dx dv)$
and the boundedness of the operators $A_1$ and $A_2$ from $L^1(O,
|v|dx dv)$ to $L^1(O,dx dv)$ and using \eqref{2.2}, we obtain that
${\mathcal R}$ is a well defined and bounded operator from $L^1(O, |v|dx
dv)$ to $L^1(F_+)$.  We shall use the following lemma for the kernel
distribution of ${\mathcal R}$.
\begin{lemma}
  \label{lem:dec1}
We have the following decomposition:
\begin{equation}
{\mathcal R}\psi(x,v)=\int_O\beta(x,v,x',v')\psi(x',v')dx'dv',\label{S0a}
\end{equation}
for a.e. $(x,v)\in F_+$ and for any $\psi\in L^1(O, |v|dx dv)$, where
\begin{equation}
0\le|v'|^{-1}\beta\in L^\infty(O,L^1(F_+)).\label{S0b}
\end{equation}

In addition if $k\in L^\infty(\R^n\times V\times V)$, then for any
$\ep'>0$, $\delta>0$, and any $1<p<1+{1\over n-1}$ there exists some
nonnegative constant $C(\ep',\delta, p)$ such that
\begin{eqnarray}
&&\left\|\int_V\int_{R\hat v+ \Pi_v(R)}
  \phi(x,v)\beta(x,v,x',v')|v|dx dv\right\|_{L^\infty(O_{x',v'})}\nonumber\\  
&&\le
 C(\ep',\delta,p)\left(\int_V\int_{R\hat v+\Pi_v(R)}
  \!\!\!\!\!\!\!\!\!  \!\!\!  
  |\phi(x,v)|^{p'}dxdv\right)^{1\over p'}
+\ep'\|\phi\|_{L^\infty(F_+)},\label{S0cb}
\end{eqnarray} 
for any continuous compactly supported function $\phi$ on $F_+$ such
that ${\rm supp}\phi\subset \{(x,v)\in F_+\ |\ |v|<\delta^{-1}\}$, and
where ${p'}^{-1}+p^{-1}=1$.
\end{lemma}
Lemma \ref{lem:dec1} is proved in Section 6. The last inequality shows
that the kernel of the second scattering operator $\mathcal R$ is more
regular than is indicated in \eqref{S0b}. When $V$ is bounded, then we
can choose $\ep'=0$ in \eqref{S0cb}, in which case we obtain that
$|v'|^{-1}\beta\in L^\infty(O,L^p(F_+))$ for $1<p<{n\over n-1}$.
This regularity is sufficient (while that described in \eqref{S0b} is
not) to show that multiple scattering contributions do not interfere
with our stability estimates.  Taking account of Lemma \ref{lem:dec1},
we have the following decomposition for the albedo operator.
\begin{lemma}
  \label{lem:dec2}
  Under condition \eqref{c3}, the following equality in the
  distributional sense is valid
  \begin{equation}
  \begin{array}{rcl}
  \A\phi_-(x,v)&=& \dint_V\int_{\Pi_{v'}(R)}\alpha(x,v,x',v')
  \phi_-(x'-R\hat{v'},v')dx'dv'\\
  &+&\dint_O\beta(x,v,x',v')((I+K)^{-1}J\phi_-)(x',v')dx'dv', \label{S1}
  \end{array}
  \end{equation}
  for a.e. $(x,v)\in F_+$ and for any $C^1$ compactly supported
  function $\phi_-$ on $F_-$, where
  \begin{eqnarray}
\alpha(x,v,x',v')&=&\alpha_1(x,v,x',v')+\alpha_2(x,v,x',v'),\label{S2}\\
\alpha_1(x,v,x',v')&=&e^{-|v|^{-1}\int_0^{2R}\sigma (x-s\hat v,v)ds }\delta_v(v')\delta_{x-(x\hat{v'})\hat{v'}}(x'),\label{S2a}\\
\alpha_2(x,v,x',v')&=&|v|^{-1}\int_0^{2R}e^{-|v|^{-1}\int_0^t\sigma(x-s\hat v,v)ds-|v'|^{-1}\int_0^{R+(x-t\hat v)\hat{v'}}\sigma(x-t\hat v-s\hat{v'},v')ds}\nonumber\\
&&\times k(x-t\hat v,v',v)\delta_{x-t\hat v-((x-t\hat v)\hat{v'})\hat{v'}}(x')dt,\label{S2b}
  \end{eqnarray}
  for a.e. $(x,v)\in F_+$ and $(x',v')\in F$,  and where $\beta$ is given by \eqref{S0a}.
\end{lemma}
Lemma \ref{lem:dec2} is proved in Section 6. The above decomposition
is similar to that obtained in \cite{choulli-stefanov-IP,CS-OSAKA-99}
except that the multiple scattering contribution is written here in
terms of the distribution kernel of $\mathcal R$ rather than that of
$\mathcal R(I+K)^{-1}J$.

\section{Stability estimates}
\label{sec:stab}
In this section, we give stability estimates for the reconstruction of
the absorption and scattering coefficient from the albedo operator
following the approach in \cite{W-AIHP-99}.

%\subsection{Additional conditions and notation}
We assume that conditions \eqref{eq:hyp1} and \eqref{c3} are satisfied
and that there exists a convex open subset $X$ of $\R^n$ with $C^1$
boundary $\pa X$ such that $\bar X\subset B(0,R):=\{x\in \R^n\ | \ 
|x|<R\}$ and
\begin{equation}
  \label{eq:hyp2}
  \begin{array}{l}
\textrm{the function }0\le\sigma_{|X\times V} \textrm{ is continous and bounded in }X\times V,\\[1mm]
\textrm{the function }0\le k_{|X\times V\times V}\textrm{ is continous and bounded in }X\times V\times V,\\[1mm]
\sigma(x,v)=k(x,v,v')=0\textrm{ for }x\not\in \bar X,\ (v,v')\in V\times V.
  \end{array}
\end{equation}
%% \begin{eqnarray}
%% &&\textrm{the function }0\le\sigma_{|X\times V} \textrm{ is continuous and bounded on }X\times V,\label{h1}\\
%% &&\textrm{the function }0\le k_{|X\times V\times V}\textrm{ is continuous and bounded on }X\times V\times V,\label{h1b}\\
%% &&\sigma(x,v)=k(x,v,v')=0\textrm{ for }x\not\in \bar X,\ (v,v')\in V\times V.\label{h1c}
%% \end{eqnarray}

Let $(\tilde \sigma, \tilde k)$ be a pair of absorption and scattering
coefficients that also satisfy \eqref{eq:hyp1}, \eqref{c3}, and
\eqref{eq:hyp2}.  We denote by a superscript $\tilde{}$ any object
(such as the albedo operator $\tilde \A$ or the distribution
kernels $\tilde \alpha_i$, $i=1,2$) associated to $(\tilde \sigma,
\tilde k)$.
 
Let $(x_0',v_0')\in F$ such that the intersection of $X$ and the
straight line $\{tv_0'+x_0'\ |\ t\in \R\}$ is not empty. The point
$(x_0'-R\hat {v_0'},v_0')\in F_-$ models the incoming condition and is
fixed in the analysis that follows.  For $\ep>0$ let $f_\ep\in
C_0^{\infty}(F_-)$ such that $\|f_\ep\|_{F_-}=1$, $f_\ep\ge 0$ and
${\rm supp} f_\ep\subset\{(x'-R\hat{v'},v')\in F_-\ | \ 
|v'-v_0'|+|x'-x_0'|<\ep\}$.  Hence $|v'|f_\ep$ is a smooth
approximation of the delta function on $F_-$ at
$(x_0'-R\hat{v_0'},v_0')$ as $\ep\to 0^+$ and is thus an admissible
incoming condition in $L^1(F_-)$. The support of $f_\ep$ is represented
in Fig. \ref{fig:1}.

For a.e. $(x,v)\in F$, $t\in \R$ and $v'\in V$ let $E(x,t,v,v')\ge 0$ be  defined by
\begin{equation}
E(x,t,v,v'):=e^{-|v|^{-1}\int_{-R}^t\sigma(x-s\hat v,v)ds-|v'|^{-1}\int_0^{R+(x-t\hat v)\hat v'}\sigma(x-t\hat v-s\hat {v'},v')ds}.\label{st2c0}
\end{equation}
Replacing $\sigma$ by $\tilde \sigma$ in \eqref{st2c0} we also define $\tilde E(x,t,v,v')$ for a.e. $(x,v)\in F$, $t\in \R$ and $v'\in V$.

Let $\delta>0$ and let $\phi$ be any compactly supported continuous
function on $F_+$ such that $\|\phi\|_\infty\le 1$ and
\begin{equation}
{\rm supp }\phi\subseteq \{(x,v)\in F_+\ |\ |v|<\delta^{-1}\}.\label{hphi}
\end{equation}  
Then using \eqref{S1} and \eqref{st2c0} we obtain for $\ep>0$ that
\begin{equation}
\int_V\int_{R\hat v +\Pi_v(R)}|v|\phi(x,v)(\A-\tilde \A)f_\ep(x,v)dxdv=I_1(\phi,\ep)+I_2(\phi,\ep)+I_3(\phi,\ep),\label{st1}
\end{equation}
where
\begin{eqnarray}
I_1(\phi,\ep)&=&\int_V\int_{\Pi_v(R)}|v|\phi(x+R\hat v,v)\left(e^{-|v|^{-1}\int_{-R}^R\sigma (x-s\hat v,v)ds }\right.\nonumber\\
&&\left.-e^{-|v|^{-1}\int_{-R}^R\tilde\sigma (x-s\hat v,v)ds }\right)f_\ep(x-R\hat v,v)dxdv\label{st2a}\\
I_2(\phi,\ep)&=&\int_{V\times V}\int_{\Pi_v(R)}\phi(x+R\hat v,v)\int_{-R}^R\left(k(x-t\hat v,v',v)E(x,t,v,v')\right.\nonumber\\
&&\left. -\tilde k(x-t\hat v,v',v)\tilde E(x,t,v,v')\right)\label{st2b}\\
&& f_\ep(x-t\hat v-(x-t\hat v)\hat {v'}-R\hat{v'},v')dt dxdvdv',\nonumber\\
I_3(\phi,\ep)&=&I_3^1(\phi,\ep)-I_3^2(\phi,\ep),\label{st2c}
\end{eqnarray}
and where
\begin{equation}
I_3^1(\phi,\ep)=\int_V\int_{R\hat v +\Pi_v(R)}
  \!\!\!\!\!\!\!\!\!\!\!\!\!\!\!\!\!\!
  |v|\phi(x,v)\int_O\beta(x,v,x',v')((I+K)^{-1}Jf_\ep)(x',v') dx'dv'dxdv,\label{st2ca}
\end{equation}
\begin{equation}
I_3^2(\phi,\ep)=\int_V\int_{R\hat v +\Pi_v(R)}
  \!\!\!\!\!\!\!\!\!\!\!\!\!\!\!\!\!\!
  |v|\phi(x,v)\int_O\tilde\beta(x,v,x',v')((I+\tilde{K})^{-1}\tilde Jf_\ep)(x',v') dx'dv'dxdv.\label{st2cb}
\end{equation}

In addition using the estimate $\|\phi\|_\infty\le 1$, item ii of
Proposition \ref{prop:albedo} and the definition of $f_\ep$, we obtain
\begin{eqnarray}
&&\left|\int_V\int_{R\hat v+\Pi_v(R)}|v|\phi(x,v)(\A-\tilde \A)f_\ep(x,v)dxdv\right|\le \|(\A-\tilde\A)f_\ep\|_{F_+}\nonumber\\
 &&
\le \|\A-\tilde\A\|_{{\mathcal L}(L^1(F_-),L^1(F_+))}\|f_\ep\|_{F_-}
=\|\A-\tilde\A\|_{{\mathcal L}(L^1(F_-),L^1(F_+))}.\label{st2d}
\end{eqnarray}

\subsection{First stability estimate}

We now prove a stability estimate under conditions \eqref{eq:hyp1},
\eqref{c3}, and \eqref{eq:hyp2}.  Taking \eqref{st2a}--\eqref{st2c}
into account, we obtain the following preparatory lemma:
\begin{lemma}
  \label{lem:limits}
  Assume that $(\sigma,k)$ and $(\tilde\sigma,\tilde k)$ satisfy
  conditions \eqref{eq:hyp1}, \eqref{c3}, and \eqref{eq:hyp2}. Then
  the following limits and estimate hold:
\begin{eqnarray}
I_1(\phi,\ep)&\underset{\ep\to 0^+}{\longrightarrow}&\phi(x_0'+R\hat {v_0'},v_0')\label{l5.2a}\\
&&\times\left(e^{-|v_0'|^{-1}\int_{-R}^R\sigma (x_0'-s\hat {v_0'},v_0')ds }
-e^{-|v_0'|^{-1}\int_{-R}^R\tilde\sigma (x_0'-s\hat {v_0'},v_0')ds }\right),\nonumber\\
I_2(\phi,\ep)&\underset{\ep\to 0^+}{\longrightarrow}&I_2^1(\phi)+I_2^2(\phi),\label{l5.2b}
\end{eqnarray}
for any compactly supported continuous function $\phi$ on $F_+$, where
\begin{eqnarray}
I_2^1(\phi)&=&{1\over |v_0'|}\int_V\int_{-R}^R(k-\tilde k)(x_0'+t'\hat{v_0'},v_0',v)\label{l5.2c}\\
&&\times\left[\phi(x+R\hat v,v)E(x,t,v,v_0')\right]_{t=t(x_0',v_0',t',v)\atop x=x(x_0',v_0',t',v)}dt'dv,
\nonumber
\end{eqnarray}
\begin{eqnarray}
I_2^2(\phi)&=&{1\over |v_0'|}\int_V\int_{-R}^R\tilde k(x_0'+t'\hat{v_0'},{v_0'},v)\label{l5.2d}\\
&&\times\left[\phi(x+R\hat v,v)(E-\tilde E)(x,t,v,v_0')\right]_{t=t(x_0',v_0',t',v)\atop x=x(x_0',v_0',t',v)}dt'dv,
\nonumber
\end{eqnarray}
where $E$ and $\tilde E$ are defined by \eqref{st2c0} and
\begin{equation}
(t(x_0',v_0',t',v),x(x_0',v_0',t',v))=\left(-(x_0'+t'\hat{v_0'})\hat{v},x_0'+t'\hat{v_0'}-((x_0'+t'\hat{v_0'})\hat{v})\hat{v}\right),\label{l5.2da}
\end{equation}
for $t'\in \R$.  In addition, for all $\ep'>0$, $\delta>0$ and for all
$1<p<1+{1\over n-1}$ there exists some nonnegative real valued
constant $C(\ep',\delta,p)$ such that
\begin{equation}
\sup_{\ep>0}|I_3(\phi,\ep)|\le C\left(C(\ep',\delta,p)\left(\int_V\int_{\Pi_v(R)}\chi_{{\rm supp}\phi}(x+R\hat{v},v)dx dv\right)^{1\over p'}+\ep'\right),\label{l5.2e}
\end{equation}
for any compactly supported continuous function $\phi$ on $F_+$, which
satisfies $\|\phi\|_\infty\le 1$ and \eqref{hphi} for $\delta>0$,
where ${p'}^{-1}+p^{-1}=1$ and
\begin{equation}
C:=2R\|(I+K)^{-1}\|_{{\mathcal L}(L^1(O,|v|dx dv))}+2R\|(I+\tilde K)^{-1}\|_{{\mathcal L}(L^1(O,|v|dx dv))}.\label{l5.2f}
\end{equation}
\end{lemma}
Lemma \ref{lem:limits} is proved in Section 5.  \vskip 2mm

Taking account of Lemma \ref{lem:limits} and \eqref{st2d}, and
choosing an appropriate sequence of functions ``$\phi$'', we obtain
the main result of the paper:
\begin{theorem}
  \label{thm:stab1}
   Assume that $n\ge 3$ and that 
  $(\sigma,k)$ and $(\tilde\sigma,\tilde k)$ satisfy conditions
  \eqref{eq:hyp1}, \eqref{c3}, and \eqref{eq:hyp2}. Then the
  following estimates are valid:
\begin{equation}
\left|e^{-|v_0'|^{-1}\int_{-R}^R\sigma (x_0'-s\hat {v_0'},v_0')ds }
-e^{-|v_0'|^{-1}\int_{-R}^R\tilde\sigma (x_0'-s\hat {v_0'},v_0')ds }\right|\le 
\|\A-\tilde\A\|_{{\mathcal L}(L^1(F_-),L^1(F_+))};\label{t5.1a}
\end{equation}
\begin{eqnarray}
&&|v_0'|^{-1}\dint_V\int_{-R}^R\left|(k-\tilde k)(x_0'+t'\hat{v_0'},v_0',v)\right|\left[E(x,t,v,v_0')\right]_{t=t(x_0',v_0',t',v)\atop x=x(x_0',v_0',t',v)}dt'dv\nonumber\\
&\le&
2R|v_0'|^{-1}\|\tilde\sigma_p(x_0'+t'\hat{v_0'},{v_0'})\|_{L^\infty(\R_{t'})}\sup\limits_{(x,v)\in F\atop t\in \R}\left|(E-\tilde E)(x,t,v,v_0')\right|\nonumber\\
&&+\|\A-\tilde\A\|_{{\mathcal L}(L^1(F_-),L^1(F_+))},\label{t5.1b}
\end{eqnarray}
where $E$ and $\tilde E$ are defined by \eqref{st2c0}, and where
$(t(x_0',v_0',t',v), x(x_0',v_0',t',v))$ is defined by \eqref{l5.2da}
for $t'\in \R$ and $v\in V$.
\end{theorem}
%{\bf Theorem 3.1.}  
%}
Theorem \ref{thm:stab1} is proved in Section 5.
\begin{remark}
  \label{rem:L} Under condition \eqref{c4}, we can obtain similar  estimates 
  to those in Theorem \ref{thm:stab1} for the albedo operator defined
  on $L(F_-)$ with values in $L(F_+)$.  Note that
  $L(F_\pm)=L^1(F_\pm)$ when $V$ is bounded.
\end{remark}
\subsection{Stability results under additional regularity assumptions}

The second inequality in Theorem \ref{thm:stab1} provides an $L^1$
stability result for $k(x,v',v)$ provided that $\sigma(x,v)$ is known.
The first inequality in Theorem \ref{thm:stab1} shows that the Radon
transform of $\sigma(x,v)$ is stably determined by the albedo
operator.  Because the inverse Radon transform is an unbounded
operation, additional constraints, including regularity constraints, on
$\sigma$ are necessary to obtain a stable reconstruction.  We assume
here that
\begin{equation} 
\big\{{v\over |v|}\ | \ v\in V\big\}=\S^{n-1},\ \V_0:=\sup_{v\in V}|v|<\infty,
\label{h3}
\end{equation}
and that the absorption coefficient $\sigma$ does not depend on the
velocity variable, i.e. $\sigma(x,v)=\sigma(x),$ $x\in \R^n$; see also
remark \ref{rem:sigma} below.  Then let
\begin{eqnarray}
&&\!\!\!\!\!{\mathcal M}:=\big\{(\sigma(x), k(x,v',v)) \in L^\infty(\R^n)\times L^\infty(\R^n\times V\times V) \ |\ (\sigma,k)\textrm{ satisfies } \eqref{eq:hyp2}, 
\nonumber\\
&&\!\!\!\!\!\eqref{eq:hyp1}\textrm{ and } \eqref{c3},\textrm{ and } \sigma_{|X}\in H^{{n\over 2}+\tilde r}(X), \|\sigma\|_{H^{{n\over 2}+\tilde r}(X)}\le M, \|\sigma_p\|_{\infty}\le M
\big\},\label{5.1}
\end{eqnarray}
for some $\tilde r>0$ and $M>0$.  Using Theorem \ref{thm:stab1} for
any $(x_0',v_0')\in F$ such that the intersection of $X$ and the straight line
$\{tv_0'+x_0'\ |\ t\in \R\}$ is not empty, we obtain
the following theorem. 
\begin{theorem}
  \label{thm:stab2} 
 Assume that $n\ge 3$. Under condition \eqref{h3}, for any $(\sigma,k)\in {\mathcal M}$ and $(\tilde \sigma, \tilde k)\in {\mathcal M}$ the following stability estimates are valid:
\begin{equation}
\|\sigma-\tilde\sigma\|_{H^s(X)}\le C_1\|\A-\tilde\A\|_{{\mathcal L}(L^1(F_-),L^1(F_+))}^\theta,\label{t5.2a}
\end{equation}
where ${-{1\over 2}}\le s<{n\over 2}+\tilde r$, $\theta={n+2(\tilde r-s)\over n+1+2\tilde r}$, and $C_1=C_1(R,X,\v_0,\V_0, M, s, \tilde r)$; 
\begin{eqnarray}
&&\int_V\int_{-R}^R\left|k(x_0'+t'\hat{v_0'},v_0',v)-\tilde k(x_0'+t'\hat{v_0'},v_0',v)\right|dt'dv\nonumber\\
&\le&
C_2\|\A-\tilde\A\|_{{\mathcal L}(L^1(F_-),L^1(F_+))}^\theta\left(1+\|\A-\tilde\A\|_{{\mathcal L}(L^1(F_-),L^1(F_+))}^{1-\theta}\right),\label{t5.2b}
\end{eqnarray}
for $(x_0', v_0')\in F$ such that $x_0'+t'v_0'\in X$ for some $t'\in \R$, and
where $\theta={2(\tilde r-r)\over n+1+2 \tilde r}$, $0< r<\tilde r$, and $C_2=C_2(R,X,\v_0,\V_0, M, r, \tilde r)$;
in addition,
\begin{eqnarray}
&&\|k-\tilde k\|_{L^1(\R^n\times V\times V)}\nonumber\\
&\le& C_3\|\A-\tilde\A\|_{{\mathcal L}(L^1(F_-),L^1(F_+))}^\theta\left(1+\|\A-\tilde\A\|_{{\mathcal L}(L^1(F_-),L^1(F_+))}^{1-\theta}\right),\label{t5.2c}
\end{eqnarray}
where $\theta={2(\tilde r-r)\over n+1+2 \tilde r}$, $0< r<\tilde r$, and $C_3=C_3(R,X,\v_0,\V_0, M, r, \tilde r)$.
\end{theorem}
%Theorem 3.2.  \vskip 2mm
Theorem \ref{thm:stab2} is proved in Section 5.

\begin{remark}
  \label{rem:sigma} 
  Theorem \ref{thm:stab2} can be extended to the case
  $\sigma=\sigma(x,|v|)$ and $V=\{v\in \R^n\ |\ 0<\lambda_1\le |v|\le
  \lambda_2<\infty\}$. In this case the class ${\mathcal M}$ is replaced
  by the class
\begin{eqnarray}
&&\!\!\!\!\!{\mathcal N}:=\big\{(\sigma(x,|v|), k(x,v',v)) \in L^\infty(\R^n\times V)\times L^\infty(\R^n\times V\times V) \ |
\nonumber\\
&&\!\!\!\!\! (\sigma,k)\textrm{ satisfies } \eqref{eq:hyp2},\ \eqref{eq:hyp1}\textrm{ and } \eqref{c3},  \|\sigma_p\|_{\infty}\le M \textrm{ and for any }\lambda\in (\lambda_1,\lambda_2),\nonumber\\
&&\!\!\!\!\!\sigma_{|X}(.,\lambda)\in H^{{n\over 2}+\tilde r}(X) , \sup_{\lambda\in (\lambda_1,\lambda_2)}\|\sigma(., \lambda)\|_{H^{{n\over 2}+\tilde r}(X)}\le M
\big\}.\label{5.1'}
\end{eqnarray}
Then the left-hand side of \eqref{t5.2a} is replaced by
$\sup_{\lambda\in (\lambda_1,\lambda_2)}\|\sigma(.,
\lambda)\|_{H^{{n\over 2}+\tilde r}(X)}$ 

\noindent whereas the right-hand side
of \eqref{t5.2a} and estimates \eqref{t5.2b}--\eqref{t5.2c} remain
unchanged (see the proof of Theorem \ref{thm:stab2}).
\end{remark}

\section{Stability in $\Gamma_\pm$}
\label{sec:stab2}

We now come back to the original geometry in \eqref{eq:tr} and present
a similar stability result (Theorem \ref{thm:stab3} below) to Theorem
\ref{thm:stab2}. The case of a scattering coefficient
$k(x,v',v)=k(v',v)$ that does not depend of the space variable $x$ was
studied in \cite{W-AIHP-99}. We now introduce the notation we need to
state our stability result.

%\subsection{New boundary data}

Recall that $X\subset \R^n$, $n\ge 2$, is an open bounded subset with
$C^1$ boundary $\pa X$, and that $V$ is $\S^{n-1}$ or an open subset
of $\R^n$ which satisfies $\v_0:=\inf_{v\in V}|v|>0$,
%Let $\Gamma_{\pm}=\{(x,v)\in \pa X\times V; \pm n(x)v>0\}$ where $n(x)$
%denotes the outward normal vector to $\pa X$ at $x\in \pa X$.
and that the linear stationary Boltzmann transport equation in
$X\times V$ takes the form
\begin{eqnarray} \label{s4.1}
&&v\nabla_xf(x,v)+\sigma(x,v)f(x,v)-\int_Vk(x,v',v)f(x,v')dv'=0\textrm{ in }X\times V,\\
&&f_{|\Gamma_-}=f_-.\nonumber
\end{eqnarray}
We assume that $(\sigma,k)$ is admissible if
\begin{equation} \label{eq:hyp1p}
  \begin{array}{l}
0\le \sigma\in L^{\infty}(X\times V),\\
k(x,v',v)\textrm{ is a measurable function on  }X\times V\times V, \textrm{ and }\\
0\le k(x,v',.)\in L^1(V)\textrm{ for a.e. }(x,v')\in X\times V \\
\sigma_p(x,v')=\int_Vk(x,v',v)dv\textrm{ belongs to } L^{\infty}(X\times V).
\end{array}
\end{equation}
For $(x,v)\in (X\times V)\cup \Gamma_\mp$, let $\tau_\pm(x,v)$ be the
real number defined by $\tau_\pm(x,v)=\sup\{t>0\ | x\pm sv \in X
\textrm{ for all }s\in (0,t) \}$. For $(x,v)\in X\times V$, let
$\tau(x,v)$ be defined by
$$
\tau (x,v)=\tau_+(x,v)+\tau_-(x,v).
$$ 
For $(x,v)\in \Gamma_\mp$, we put $\tau(x,v)=\tau_\pm(x,v)$.
We consider the measure $d\xi(x,v)=|n(x)v|d\mu(x)dv$  on $\Gamma_{\pm}$. 
%Our sources will be placed on $\Gamma_-$ and  we shall take measurements on $\Gamma_+$.
%\subsection{The albedo operator}
We still use the notation $T_0$, $T_1$, $T$, $A_1$, and $A_2$ as in
\eqref{eq:opTA}
%% We shall keep the following notations 
%% \begin{eqnarray*}
%% && T_0f=-v\nabla_x f\ \textrm{in the distributional sense},\ \ A_1f=-\sigma f,\ A_2f=\int_Vk(x,v',v)f(x,v')dv',\\
%% &&T_1=T_0+A_1,\ T=T_0+A_1+A_2=T_1+A_2.
%% \end{eqnarray*}
and introduce the following Banach space
\begin{eqnarray*}
&&{\rm W}:=\{f\in L^1(X\times V); T_0f\in L^1(X\times V),\ \tau^{-1}f\in L^1(X\times V)\},\\
&&\|f\|_{{\rm W}}=\|T_0f\|_{L^1(X\times V)}+\|\tau^{-1}f\|_{L^1(X\times V)}.
\end{eqnarray*}
We recall the following  trace formula (see Theorem 2.1 of \cite{CS-OSAKA-99})
\begin{equation}
\|f_{|\Gamma_\pm}\|_{L^1(\Gamma_\pm,d\xi)}\le \|f\|_{\rm W}, \textrm{ for }f\in {\rm W}.\label{s4.5}
\end{equation} 
Estimate \eqref{s4.5} is the analog of the estimate
\eqref{2.2} in the previous measurement setting.  For a continuous
function $f_-$ on $\Gamma_-$, we define ${\mathcal J}f_-$ as the extension
of $f_-$ in $X\times V$ given by :
\begin{equation}
{\mathcal J}f_-(x,v)=e^{-\int_0^{\tau_-(x,v)}\sigma (x-sv,v)ds }f_-(x-\tau_-(x,v) v,v),\ (x,v)\in X\times V.\label{s4.6}
\end{equation}
Note that ${\mathcal J}$ has the following trace property (see
Proposition 2.1 of \cite{CS-OSAKA-99}):
\begin{equation}
\|{\mathcal J}f_-\|_{\rm W}\le C\|f_-\|_{L^1(\Gamma_-,d\xi)},\label{s4.7}
\end{equation}
for  $f_-\in L^1(\Gamma_-,d\xi)$, where $C=1+{\rm diam}(X)\v_0^{-1}\|\sigma\|_\infty$ and where ${\rm diam}(X):=$ 

\noindent $\sup_{x,y\in X}|x-y|$.
Estimate \eqref{s4.7} is the analog of estimate \eqref{2.4} in the previous measurement setting.

\subsection{Existence theory for the albedo operator}

We denote by ${\mathcal K}$ the bounded operator of $L^1(X\times V, \tau^{-1}dx dv)$ defined by
$$
{\mathcal K}f(x,v)=-\int_0^{\tau_-(x,v)}e^{-\int_0^t\sigma(x-sv,v)ds}\int_Vk(x,v',v)f(x-tv,v')dv'dt,\ (x,v)\in X\times V.
$$
for all $f\in L^1(X\times V, \tau^{-1}dx dv)$.  We transform the
stationary linear Boltzmann transport equation \eqref{s4.1} into the
following integral equation
\begin{equation}
(I+{\mathcal K})f={\mathcal J}f_-.\label{s4.8}
\end{equation}

We have the following proposition, which is the analog of Proposition
\ref{prop:albedo}.
\begin{proposition}
  \label{prop:fwd2}
Assume that
\begin{eqnarray}
&&\textrm{the bounded operator }I+{\mathcal K} \textrm{ in } L^1(X\times V,\tau^{-1}dx dv) \textrm{ admits a bounded inverse}\nonumber\\
&&\textrm{in } L^1(X\times V,\tau^{-1}dx dv).\label{c3'} 
\end{eqnarray}
Then
\begin{itemize}

\item[i.]the integral equation \eqref{s4.8} is uniquely solvable for all $f_-\in L^1(\Gamma_-,d\xi)$,  and  $f\in {\rm W}$,
  
\item[ii.]the operator ${\rm A}: L^1(\Gamma_-,d\xi)\to
  L^1(\Gamma_+,d\xi)$, $f_-\to f_{|\Gamma_+}$, is a bounded operator.
  This operator is called the albedo operator ${\rm A}$ for
  \eqref{s4.1}.

\end{itemize}
\end{proposition}
The above proposition can be proved by slightly modifying the proofs
of Propositions 2.3 and 2.4 of \cite{CS-OSAKA-99}.
\begin{remark}
  \label{rem:var}
\begin{itemize}
\item[i.] Assume that $X$ is also convex. Let $f\in L^1(F_\pm)$ be
  such that ${\rm supp}f\subseteq\{(x,v)\in F_\pm \ |\ x+tv\in X
  \textrm{ for some }t\in \R\}$, where $F_\pm$ is defined by
  \eqref{0.3} and $R>{\rm diam}(X)$. Then we obtain that:
\begin{equation}
\int_V\int_{\Pi_v(R)}f(x\pm R\hat v,v)|v|dxdv=\int_{\Gamma_\pm}f(\gamma_\pm(x,v),v)d\xi(x,v),\label{s4.9}
\end{equation}
where $\gamma_\pm(x,v)=x-(x\hat v)\hat v\pm R\hat v$ for any $(x,v)\in
\Gamma_\pm$.  Therefore, considering results on existence of the
albedo operator ${\rm A}$ obtained in \cite{CS-OSAKA-99} and our
assumptions \eqref{eq:hyp1}, equality \eqref{s4.9} leads us to define
the albedo operator $\A$ from $L^1(F_-)$ to $L^1(F_+)$.
\item[ii.]The condition \eqref{c3'} is satisfied 
under either of the following constraints:
\begin{eqnarray}
  \|\tau\sigma_p\|_\infty&<&1,\label{c1'}\\
  \sigma-\sigma_p&\ge& 0.
\end{eqnarray}
\item[iii.]Assume that
\begin{eqnarray}
&&\textrm{the bounded operator }I+{\mathcal K} \textrm{ in } L^1(X\times V) \textrm{ admits a bounded inverse}\nonumber\\
&&\textrm{in } L^1(X\times V).\label{c4'} 
\end{eqnarray}
Then we can define the albedo operator from $L^1(\Gamma_-,
d\tilde\xi)$ to $L^1(\Gamma_+,d\tilde\xi)$ where $d\tilde
\xi=\min(\tau, \lambda) d\xi$ and where $\lambda$ is a positive
constant. To prove this latter statement, we need trace results for
the functions $f\in \tilde {\rm W}:=\{f\in L^1(X\times V)\ |\ T_0f\in
L^1(X\times V)\}$.
\item[iv.]Under \eqref{c1'} and the condition $\|\tau
  \sigma\|<\infty$, the existence of the albedo operator ${\rm
    A}:L^1(\Gamma_-,d\xi)\to L^1(\Gamma_+,d\xi)$ is proved in
  \cite{CS-OSAKA-99} (Proposition 2.3) when $V$ is an open subset of
  $\R^n$ (the condition $\inf_{v\in V}|v|>0$ is not required).
\item[v.]Under the condition $\sigma-\sigma_p\ge \nu>0$, the existence
  of the albedo operator ${\rm A}:L^1(\Gamma_-,d\tilde \xi)\to
  L^1(\Gamma_+,d\tilde \xi)$ is proved in \cite{CS-OSAKA-99}
  (Proposition 2.4) when $V$ is an open subset of $\R^n$ (the
  condition $\inf_{v\in V}|v|>0$ is not required).
\end{itemize}
\end{remark}
Finally under \eqref{c3'}, we also obtain a decomposition of the
albedo operator ${\rm A}$ similar to that of $\A$ given in Lemma
\ref{lem:dec2}.

\subsection{Stability estimates}
We assume that $X$ is convex and
\begin{equation}
  \label{eq:hyp2p}
  \begin{array}{l}
\textrm{the function }0\le\sigma \textrm{ is continous and bounded on }X\times V,
\\
\textrm{the function }0\le k\textrm{ is continous and bounded on }X\times V\times V.
  \end{array}
\end{equation}
%% \begin{eqnarray}
%% &&\textrm{the function }\sigma \textrm{ is continuous and bounded on }X\times V,\label{h1'}\\
%% &&\textrm{the function }k\textrm{ is continuous and bounded on }X\times V\times V.\label{h1b'}
%% \end{eqnarray}

Let $(\tilde \sigma, \tilde k)$ be a pair of absorption and scattering
coefficients that also satisfy \eqref{eq:hyp2p}, \eqref{eq:hyp1p} and
\eqref{c3'}.  Let $\tilde {\rm A}$ be the albedo operator from
$L^1(\Gamma_-,d\xi)$ to $L^1(\Gamma_+,d\xi)$ associated to $(\tilde
\sigma,\tilde k)$. We can now obtain stability results similar to
those in Lemma \ref{lem:limits} and Theorem \ref{thm:stab1}.  Consider
\begin{eqnarray}
{\rm M}&:=&\big\{(\sigma(x), k(x,v',v)) \in H^{{n\over 2}+\tilde r}(X)\times C(X\times V\times V) \ | 
\nonumber\\
&&(\sigma,k)\textrm{ satisfies } \eqref{eq:hyp2p}, \ \eqref{c3'}, \|\sigma\|_{H^{{n\over 2}+\tilde r}(X)}\le M, \|\sigma_p\|_{\infty}\le M
\big\}\label{s4.10}
\end{eqnarray}
for some $\tilde r>0$ and $M>0$.  We obtain the following theorem.
\begin{theorem}
  \label{thm:stab3}
  Assume $n\ge 3$. Under conditions \eqref{h3}, for any $(\sigma,k)\in
  {\rm M}$ and $(\tilde \sigma, \tilde k)\in {\rm M}$, the following
  stability estimates are valid:
  \begin{equation}
\|\sigma-\tilde\sigma\|_{H^s(X)}\le C_1\|{\rm A}-\tilde{\rm A}\|_{{\mathcal L}(L^1(\Gamma_-,d\xi),L^1(\Gamma_+,d\xi))}^\theta,\label{t4.1a}
  \end{equation}
  where ${-{1\over 2}}\le s<{n\over 2}+\tilde r$, $\theta={n+2(\tilde
    r-s)\over n+1+2\tilde r}$, and $C_1=C_1(X,\v_0,\V_0, M, s, \tilde
  r)$;
  \begin{eqnarray}
&&\int_V\int_0^{\tau_+(x_0',v_0')}\left|k(x_0'+t'\hat{v_0'},v_0',v)-\tilde k(x_0'+t'\hat{v_0'},v_0',v)\right|dt'dv\label{t4.1b}\\
&\le&
C_2\|{\rm A}-\tilde{\rm A}\|_{{\mathcal L}(L^1(\Gamma_-,d\xi),L^1(\Gamma_+,d\xi))}^\theta \left(1+\|{\rm A}-\tilde{\rm A}\|_{{\mathcal L}(L^1(\Gamma_-,d\xi),L^1(\Gamma_+,d\xi))}^{1-\theta}\right),\nonumber
  \end{eqnarray}
for $(x_0', v_0')\in \Gamma_-$ and where $\theta={2(\tilde r-r)\over
    n+1+2 \tilde r}$, $0< r<\tilde r$, and $C_2=C_2(X,\v_0,\V_0, M, r,
  \tilde r)$. As a consequence, we have
  \begin{eqnarray}
&&\|k-\tilde k\|_{L^1(X\times V\times V)}\nonumber\\
&\le& C_3\|{\rm A}-\tilde{\rm A}\|_{{\mathcal L}(L^1(\Gamma_-,d\xi),L^1(\Gamma_+,d\xi))}^\theta\left(1+\|{\rm A}-\tilde{\rm A}
\|_{{\mathcal L}(L^1(\Gamma_-,d\xi),L^1(\Gamma_+,d\xi))}^{1-\theta}\right),
\qquad\label{t4.1c}
  \end{eqnarray}
where $\theta={2(\tilde r-r)\over n+1+2 \tilde r}$, $0< r<\tilde r$,
and $C_3=C_3(X,\v_0,\V_0, M, r, \tilde r)$.
\end{theorem}
%{\bf Theorem 4.1.}
The proof of Theorem \ref{thm:stab3} is similar to that of Theorem
\ref{thm:stab2}.
\begin{remark}
  \label{rem:extp} 
  Theorem \ref{thm:stab3} can also be extended to the case
  $\sigma=\sigma(x,|v|)$ and $V=\{v\in \R^n\ |\ 0<\lambda_1\le |v|\le
  \lambda_2<\infty\}$. In this case the class ${\rm M}$ is replaced by
  the class
\begin{eqnarray}
{\rm N}&:=&\big\{(\sigma(x,|v|), k(x,v',v)) \in C(X\times V)\times C(X\times V\times V) \ |
\nonumber\\
&&(\sigma,k)\textrm{ satisfies } \eqref{eq:hyp2p},\ \eqref{c3'},\ \|\sigma_p\|_{\infty}\le M,\  \textrm{ and for any }\lambda\in (\lambda_1,\lambda_2),\nonumber \\
&&\sigma_{|X}(.,\lambda)\in H^{{n\over 2}+\tilde r}(X),\ \sup_{\lambda\in (\lambda_1,\lambda_2)}\|\sigma(., \lambda)\|_{H^{{n\over 2}+\tilde r}(X)}\le M
\big\}.\label{s4.10''}
\end{eqnarray}
Then the left-hand side of \eqref{t4.1a} is replaced by $\sup_{\lambda\in (\lambda_1,\lambda_2)}\|\sigma(., \lambda)\|_{H^{{n\over 2}+\tilde r}(X)}$ 

\noindent whereas the right-hand side of \eqref{t4.1a} and  estimates \eqref{t4.1b}--\eqref{t4.1c} remain 

\noindent unchanged. 
\end{remark}

%%%%%%%%%%%%%%%%%%%%%%%%%%%%%%%%%%%%%%%%%%%%%%%%%%%%%%%%%%%%%%%%%%%%%%%%%%%%%%%%%%%%%%%%%%%%%%%%%%%%%%%%%

\section{Proof of the stability results}
\label{sec:proof3}
We now prove Lemma 3.1 and Theorems 3.1 and 3.2.
\begin{proof}[Proof of Lemma \ref{lem:limits}.]
  Using the fact that $X$ is a convex subset of $\R^n$ with $C^1$
  boundary and using \eqref{eq:hyp2}, we obtain that
\begin{eqnarray} 
&&\textrm{the function }\R\times \R^n\times V\ni (t,x,v)\to \int_{-R}^t\sigma(x-s\hat v,v)ds\textrm{ is continous}\nonumber\\
&&\textrm{at any point }(\bar t,\bar x,\bar v)\textrm{ such that }\bar x+\eta\bar v\in X\textrm{ for some real }\eta.\label{stt1}
\end{eqnarray}
The same statement holds by replacing $\sigma$ by $\tilde \sigma$.
From \eqref{st2a}, it follows that
\begin{equation}
I_1(\phi,\ep)=\int_V\int_{xv=0\atop |x|<R}\Phi_1(x,v)f_{\ep}(x-R\hat v,v)|v|dx dv,\label{stt2}
\end{equation}
where $\Phi_1$ is the bounded function on $F$ defined for $(x,v)\in F$ by
\begin{eqnarray}
\Phi_1(x,v)=\phi(x+R\hat v,v)\left(e^{-|v|^{-1}\int_{-R}^R\sigma (x-s\hat v,v)ds }-e^{-|v|^{-1}\int_{-R}^R\tilde\sigma (x-s\hat v,v)ds}\right).\label{stt3}
\end{eqnarray}
%for $(x,v)\in F$.

From \eqref{stt1} and the continuity of $\phi$, it follows that
$\Phi_1$ is continuous at the point $(x_0',v_0')$ in $F$. Therefore
using \eqref{stt2} and the definition of the functions $f_{\ep}$, we
obtain $\lim_{\ep\to 0^+}I_1(\phi,\ep)=\Phi_1(x_0',v_0')$, which
implies \eqref{l5.2a}.  Performing the change of variables
$x-t\hat{v}=x'+t'\hat{v'}$ with $x'v'=0$ (``$dtdx=dt'dx'$'') in
formula \eqref{st2b} and using \eqref{eq:hyp2}, we obtain
\begin{equation}
I_2(\phi,\ep)=\int_{-R}^R\int_V\left(\int_V\int_{\Pi_{v'}(R)}\!\!\!\!\!\!\Phi_{2,t',v}(x',v')f_\ep(x'-R\hat{v'},v')|v'|dx'dv'\right)dv dt',\label{st4d}
\end{equation}
where
\begin{equation}
\Phi_{2,t',v}(x',v')=
0\textrm{ if }x'+t'\hat{v'}\not\in X,\label{st2bb}
\end{equation}
and
\begin{eqnarray}
&&\Phi_{2,t',v}(x',v')={1\over |v'|}\left(k(x'+t'\hat{v'},v',v)\left[\phi(x+R\hat v,v)E(x,t,v,v')\right]_{t=t(x',v',t',v)\atop x=x(x',v',t',v)}\right.\nonumber\\
&&\qquad\qquad-\left.\tilde k(x'+t'\hat{v'},v',v)\left[\phi(x+R\hat v,v)\tilde E(x,t,v,v')\right]_{t=t(x',v',t',v)\atop x=x(x',v',t',v)}\right)\label{sti2b}\\
&&\textrm{ if }x'+t'\hat{v'}\in X,
\nonumber
\end{eqnarray}
for $(t',v)\in (-R,R)\times V$, where
\begin{equation}
(t(x',v',t',v),x(x',v',t',v)):=\left(-(x'+t'\hat{v'})\hat v,x'+t'\hat{v'}-(x'+t'\hat{v'})\hat v\right),\label{stii2}
\end{equation}
for $x'\in \R^n$, $v,v'\in V$, $t'\in \R$.
Let $t'\in (-R,R)$ such that $x_0'+t'v_0'\in X$, and let $v\in V$. Then from \eqref{stt1}, \eqref{eq:hyp2}--\eqref{st2c0} and  \eqref{sti2b}--\eqref{stii2} it follows that $\Phi_{2,t',v}$ is continuous at the point $(x_0',v_0')$. Hence 
\begin{equation}
\int_V\int_{x'v'=0\atop |x'|<R}\Phi_{2,t',v}(x',v')f_\ep(x'-R\hat{v'},v')|v'|dx'dv'\to\Phi_{2,t',v}(x_0',v_0'),\textrm{ as }\ep\to 0^+.\label{stt2c}
\end{equation}

Moreover using \eqref{st2bb}--\eqref{sti2b} and using the estimate $\sigma\ge 0$ (and \eqref{st2c0}) and the equality $\|f_\ep\|_{F_-}=1$ we obtain
\begin{equation}
\left|\int_V\!\!\int_{x'v'=0\atop |x'|<R}\!\!\!\!\!\!\!\!\Phi_{2,t',v}(x',v')f_\ep(x'-R\hat{v'},v')|v'|dx'dv'\right|\le{\|k+\tilde k\|_\infty\|\phi\|_{L^\infty(F_+)}\chi_{{\rm supp}_V\phi}(v)
\over \v_0},
\label{stt2d}
\end{equation}
for $(t',v)\in (-R,R)\times V$, where ${\rm supp}_V\phi=\overline{\{v\in V\ |\ \exists x\in \R^n,\ x\hat v=R,\ \phi(x,v)\not=0\}}$.
From \eqref{stt2c}, \eqref{stt2d}, \eqref{st4d} it follows that $\lim_{\ep\to0^+}I_2(\phi,\ep)=\int_{-R}^R\int_V\Phi_{2,t',v'}(x_0',v_0')dv'$ $dt'$, which implies \eqref{l5.2b}.

It remains to prove \eqref{l5.2e}.  We first estimate
$\sup_{\ep>0}|I_3^1(\phi,\ep)|$.  Using \eqref{c3}, the estimate
$\|Jf_\ep\|_{L^1(O,|v|dx dv)}\le 2R\|f_\ep\|_{F_-}$ and the equality
$\|f_\ep\|_{F_-}=1$, we obtain
\begin{equation}
\|(I+K)^{-1}Jf_\ep\|_{L^1(O,|v|dx dv)}\le 2R\|(I+K)^{-1}\|_{{\mathcal L}(L^1(O,|v|dx dv))},\ \ep>0.\label{st2e}
\end{equation}

Let $\ep'>0$ and $1<p<1+{1\over n-1}$ and $p^{-1}+{p'}^{-1}=1$.  Using
\eqref{S0b}, \eqref{st2ca}, \eqref{S0cb} and the estimate
$\|\phi\|_{L^\infty(F_+)}\le 1$ we obtain
\begin{equation*}
|I_3^1(\phi,\ep)|=\left|\int_O((I+K)^{-1}Jf_\ep)(x',v') \int_V\int_{R\hat v+\Pi_v(R)}\!\!\!\!\!\!\!\!\!\!\!\!\!\!\!\!\!\!\!\phi(x,v)\beta(x,v,x',v')|v|dx dv dx' dv'\right|
\end{equation*}
\begin{eqnarray}
&&\!\!\!\!\!\!\!\le\|(I+K)^{-1}Jf_\ep\|_{L^1(O,|v|dx dv)}\left\|\int_V\int_{R\hat v+\Pi_v(R)}\!\!\!\!\!\!\!\!\phi(x,v)|v|{1\over |v'|}\beta(x,v,x',v')dx dv\right\|_{L^\infty(O)}\nonumber\\
&&\!\!\!\!\!\!\!\le{1\over \v_0}C_1\left(C(\ep',\delta,p)\left(\int_V\int_{R\hat v+\Pi_v(R)}\chi_{{\rm supp}\phi}(x,v)dx dv\right)^{1\over p'}+\ep'\right),\hfill\label{st3a}
\end{eqnarray}
where $C(\ep',\delta,p)$ is the constant from \eqref{S0cb}, and 
\begin{equation}
C_1=2R\|(I+K)^{-1}\|_{{\mathcal L}(L^1(O,|v|dx dv))}.\label{st3b}
\end{equation}
Replacing $K$, $J$, $\sigma$ and $\beta$ by $\tilde K$, $\tilde J$, $\tilde \sigma$ and $\tilde \beta$ in \eqref{st3a}--\eqref{st3b}, we obtain an estimate for 
$\sup_{\ep>0}I_3^2(\phi,\ep)$. Combining these estimates with \eqref{st2c}, we obtain \eqref{l5.2e}.
\end{proof}

\begin{proof}[Proof of Theorem \ref{thm:stab1}.]
Let $\ep_1>0$ and let $\phi_{\ep_1}$ be any compactly supported continuous function on $F_+$ which satisfies $0\le \phi_{\ep_1}\le 1$ and 
\begin{eqnarray} 
&&\phi_{\ep_1}(x+R\hat v,v)=1 \textrm{ for }(x,v)\in F,\ |x-x_0'|+|v-v_0'|<{\ep_1\over 2},\label{st5a}\\
&&{\rm supp}\phi_{\ep_1}\subseteq\{(x,v)\in F_+,\ |x-R\hat v-x_0'|+|v-v_0'|<\ep_1\}\label{st5b}.
\end{eqnarray}
From \eqref{l5.2a} and \eqref{st5a} it follows that
\begin{equation}
\lim_{\ep_1\to 0^+}\lim_{\ep\to 0^+}I_1(\ep,\phi_{\ep_1})=e^{-|v_0'|^{-1}\int_{-R}^R\sigma (x_0'-s\hat {v_0'},v_0')ds }
-e^{-|v_0'|^{-1}\int_{-R}^R\tilde\sigma (x_0'-s\hat {v_0'},v_0')ds }.\label{st6a}
\end{equation}
From \eqref{l5.2e} and \eqref{st5b} it follows that
\begin{equation}
\lim_{\ep_1\to 0^+}\limsup_{\ep\to 0^+}|I_3(\ep,\phi_{\ep_1})|=0.\label{st6b}
\end{equation}
From \eqref{l5.2c}, \eqref{l5.2d}, it follows that
\begin{equation}
|I_2^1(\phi_{\ep_1})+I_2^2(\phi_{\ep_1})|\le {2R\over \v_0}(\|k\|_\infty+\|\tilde k\|_{\infty})\int_V\chi_{{\rm supp}_V\phi_{\ep_1}}(v)dv,\label{st6c}
\end{equation}
where ${\rm supp}_V\phi_{\ep_1}=\overline{\{v\in V\ |\ \exists x\in \R^n,\ x\hat v=R,\ \phi_{\ep_1}(x,v)\not=0 \}}$.
Note that using \eqref{st5b}, we obtain
\begin{equation}
\int_V\chi_{{\rm supp}_V\phi_{\ep_1}}(v)dx dv\le \int_{v\in V\atop |v-v_0'|<\ep_1}dv\to 0,\textrm{ as }\ep_1\to 0^+.\label{st6d}
\end{equation}
From \eqref{st6d} and \eqref{st6c} it follows that
\begin{equation}
|I_2^1(\phi_{\ep_1})+I_2^2(\phi_{\ep_1})|\to 0,\textrm{ as }\ep_1\to 0^+.\label{st6e}
\end{equation}
Note also that from \eqref{st1} and \eqref{st2d}, it follows that
\begin{equation}
|I_1(\phi_{\ep_1},\ep)|\le \|{\mathcal A}-\tilde {\mathcal A}\|_{{\mathcal L}(L^1(F_-),L^1(F_+))}+|I_2^1(\phi_{\ep_1},\ep)+I_2^2(\phi_{\ep_1},\ep)|+|I_3(\phi_{\ep_1},\ep)|,\label{st6ee}
\end{equation}
for $\ep>0$ and $\ep_1>0$.
Combining \eqref{st6ee} (with ``$\phi$''$=\phi_{\ep_1}$), \eqref{st6a}, \eqref{st6b} and \eqref{st6e}, we obtain \eqref{t5.1a}. This provides us
with a stability result for the absorption coefficient.

It remains to obtain a stability result for the scattering
coefficient.  We first construct an appropriate set of functions
``$\phi$'' (see \eqref{st7h} below).  The objective is to construct a
sequence of such (smooth) functions whose support converges to the
line in $F_+$ where single scattering is restricted; see
Fig.\ref{fig:1}.  Moreover, we want these functions to be good
approximations of the sign of $k-\tilde k$ on that support. This is
the main new ingredient that allows us to obtain stability for
spatially dependent scattering coefficients.  More precisely, let
$U:=\{(t',v)\in \R\times V\ |\ x_0'+t'\hat{v_0'}\in X\textrm{ and
}(k-\tilde k)(x_0'+t'\hat{v_0'},v_0',v)>0\}$. Using \eqref{eq:hyp2}, it
follows that $U$ is an open subset of $\R \times V$. Let $(K_m)$ a
sequence of compact sets such that $\bigcup_{m\in \N}K_m=U$ and
$K_m\subseteq K_{m+1}$ for $m\in \N$.  For $m\in \N$ let $\chi_m\in
C^\infty(\R\times V,\R)$ such that $\chi_{K_m}\le \chi_m\le \chi_U$,
and let
\begin{equation}
\rho_m=2\chi_m-1.\label{st7a}
\end{equation}
Thus using \eqref{eq:hyp2} we obtain
\begin{equation}
\lim_{m\to +\infty}(k-\tilde k)(x_0'+t'\hat{v_0'},v_0',v)\rho_m(t',v)=
|k-\tilde k|(x_0'+t'\hat{v_0'},v_0',v),\label{st7u1}
\end{equation}
for $v\in V$ and $t'\in \R$ such that  $x_0'+t'v_0'\in X\cup (\R^n\backslash \bar X)$.
For $(x,v)\in F_+$ such that $v$ and $v_0'$ are linearly independent, we define 
\begin{equation}
{\rm d}(x,v):=\left|x-x_0'-((x-x_0')\hat v)\hat v-((x-x_0'){\hat v-(\hat{v_0'}\hat v)\hat{v_0'}\over\sqrt{1-(\hat{v_0'}\hat{v})^2}})
{\hat{v}-(\hat{v_0'}\hat{v})\hat{v_0'}\over\sqrt{1-(\hat{v_0'}\hat{v})^2}}\right|.\label{st7b}
\end{equation} 
For $(x,v)\in F_+$ such that $v$ and $v_0'$ are linearly independent,
we verify that ${\rm d}(x,v)=\inf_{t,t'\in \R}|x+t\hat
v-(x_0'+t'\hat{v_0'})|$ and the infimum is reached at
\begin{equation}
  \label{st7d}
  (t,t') = \Big({(x_0'-x)(\hat v-(\hat v\hat{v_0'})\hat{v_0'})\over 1-(\hat v\hat{v_0'})^2},{(x-x_0')(\hat{v_0'}-(\hat v\hat{v_0'})\hat v)\over 1-(\hat v\hat{v_0'})^2}\Big).
\end{equation}
%% \begin{eqnarray}
%% s&=&{(x_0'-x)(\hat v-(\hat v\hat{v_0'})\hat{v_0'})\over 1-(\hat v\hat{v_0'})^2},\label{st7c}\\
%% t&=&{(x-x_0')(\hat{v_0'}-(\hat v\hat{v_0'})\hat v)\over 1-(\hat v\hat{v_0'})^2}.\label{st7d}
%% \end{eqnarray}
Consider
\begin{equation}
{\mathcal V}_{\delta,l}:=  \{(x,v)\in F_+\ |\ |x-R\hat v|<R-\delta,\ | \hat{v}-{\hat{v}v_0'\over {v_0'}^2}v_0'|>\delta,\ |v|<{1\over\delta},\ {\rm d}(x,v)<{1\over l}\},\quad\label{st7e}
\end{equation}
\begin{eqnarray}
\tilde{\mathcal V}_{\delta,l}&:=&
  \{(x,v)\in F_+\ |\ |x-R\hat v|\le R-\delta-{1\over l},\ |\hat v-{\hat vv_0'\over{v_0'}^2}v_0'|\ge\delta+{1\over l},\nonumber\\
&&|v|\le \delta^{-1}-l^{-1},\ {\rm d}(x,v)\le{1\over 2l}\},\label{st7f}
\end{eqnarray}
for $0<\delta<\min(R,\v_0^{-1})$ and $l\in \N$, $l>(R-\delta)^{-1}+\delta$. For $0<\delta<\min(R,\v_0^{-1})$ and $l\in \N$, $l>(R-\delta)^{-1}+\delta$, let $\chi_{\delta,l}\in C^{\infty}_0(F_+)$ be such that
\begin{equation}
\chi_{\tilde{\mathcal V}_{\delta,l}}\le \chi_{\delta,l}\le \chi_{{\mathcal V}_{\delta,l}}.\label{st7g}
\end{equation}
Finally, for $0<\delta<\min(R,\v_0^{-1})$ and $m,l\in \N$,
$l>(R-\delta)^{-1}+\delta$, let $\phi_{\delta,m,l}\in C^\infty_0(F_+)$
be defined by
\begin{equation}
\phi_{\delta,m,l}(x,v):=\chi_{\delta,l}(x,v)\rho_m\left(t',v\right)_{\big|t'={(x-x_0')(\hat{v_0'}-(\hat v\hat{v_0'})\hat v)\over 1-(\hat v\hat{v_0'})^2}}.\label{st7h}
\end{equation}
(See \eqref{st7u1}, \eqref{st7d} and \eqref{st7ra}--\eqref{st7rb}
given below.)  Note that from \eqref{st7h} and \eqref{st7g} it follows
that
\begin{equation}
{\rm supp}\phi_{\delta,m,l}\subseteq  {\mathcal V}_{\delta,l}.\label{st7i}
\end{equation}

Using \eqref{l5.2a}, \eqref{st7i} and \eqref{st7e}, it follows that
\begin{equation}
\lim_{\ep\to 0^+}I_1(\phi_{\delta,m,l},\ep)=0,\label{st7j}
\end{equation}
for $0<\delta<\min(R,\v_0^{-1})$ and $m,l\in \N$, $l>(R-\delta)^{-1}+\delta$ (we used that $(x_0'+R\hat v_0',v_0')\not\in {\mathcal V}_{\delta,l}$).

Note that using \eqref{st7e} we obtain 
\begin{eqnarray}
\chi_{{\mathcal V}_{\delta,l}}(x,v)&\le&\chi_G(x,v)\le 1,\ l\in \N,\ l>(R-\delta)^{-1}+\delta,\label{st7ka}\\
\lim_{l\to\infty}\chi_{{\mathcal V}_{\delta,l}}(x,v)&=&\chi_{{\mathcal V}_\delta}(x,v), \label{st7kb}
\end{eqnarray}
for $(x,v)\in F_+$ and $0<\delta<\min(R,\v_0^{-1})$, where $G$ is the
compact subset of $F_+$ given by $G=\{(x,v)\in F_+\ |\ |x-R\hat v|\le
R-\delta,\ | \hat{v}-{\hat{v}v_0'\over {v_0'}^2}v_0'|\ge \delta,\ 
|v|\le \delta^{-1}\}$ and
\begin{equation}
{\mathcal V}_\delta:=\big\{(x,v)\in F_+\ |\ |x-R\hat v|<R-\delta,\ |\hat{v}-{\hat{v}v_0'\over {v_0'}^2}v_0'|>\delta,\ |v|<\delta^{-1},\ {\rm d}(x,v)=0\big\}.\label{st7l}
\end{equation}
Note also that, as $n\ge 3$, we obtain
\begin{equation}
\int_V\int_{\Pi_v(R)}\chi_{{\mathcal V}_\delta}(x+R\hat v,v)dx dv=0, \ 0<\delta<\min(R, \v_0^{-1}).\label{st7m}
\end{equation}

From \eqref{l5.2e}, \eqref{st7i}, \eqref{st7ka}--\eqref{st7m}, it follows that 
\begin{equation}
\limsup_{l\to \infty}\limsup_{\ep\to 0^+}|I_3(\phi_{\delta,m,l},\ep)|\le \ep' \textrm{ for any }\ep'>0.\label{st7qb}
\end{equation}
Hence
\begin{equation}
\lim_{l\to \infty}\limsup_{\ep\to 0^+}|I_3(\phi_{\delta,m,l},\ep)|= 0.\label{st7q}
\end{equation}

Let $0<\delta<\min(R,\v_0^{-1})$ and $m\in \N$.
Using \eqref{st7h}, \eqref{st7d} and \eqref{l5.2c}, we obtain
\begin{equation}
I_2^1(\phi_{\delta,m,l})=\int_V\int_{-R}^Rf_{\delta,m,l}(t',v)dt'dv,
\label{st7ra}
\end{equation}
for $l>(R-\delta)^{-1}+\delta$ where
\begin{eqnarray}
f_{\delta,m,l}(t',v)&:=&{1\over |v_0'|}(k-\tilde k)(x_0'+t'\hat{v_0'},v_0',v)\rho_m(t',v)\nonumber\\
&&\times\left[\chi_{\delta,l}(x+R\hat v,v)E(x,t,v,v_0')\right]_{t=t(x_0',v_0',t',v)
\atop x=x(x_0',v_0',t',v)},\label{st7rb}
\end{eqnarray}
for $t'\in \R$ and $v\in V$ such that $v$ and $v_0'$ are linearly independent, where $(t(x_0',v_0',$ $t',v),x(x_0',v_0',t',v))$ is defined by \eqref{l5.2da} for $v\in V$ and $t'\in \R$.

Using  the estimates $\sigma\ge 0,$ $|v|^{-1}\le \v_0^{-1}$, $0\le \chi_{\delta,l}\le 1$, we obtain
\begin{equation} 
|f_{\delta,m,l}(t',v)|\le {1\over \v_0}(k+\tilde k)(x_0'+t'\hat{v_0'},v_0',v)), \label{st7s}
\end{equation}
for $l>(R-\delta)^{-1}+\delta$, $t'\in \R$ and $v\in V$ such that $v$
and $v_0'$ are linearly independent. Using \eqref{eq:hyp1} and
\eqref{eq:hyp2}, we obtain that the function arising on the
right-hand side of \eqref{st7s} is integrable on $V\times (-R,R)$.  In
addition from \eqref{st7e}--\eqref{st7g}, \eqref{st7kb} and
\eqref{st7l}, it follows that
\begin{equation}
f_{\delta,m,l}(t',v)\to f_{\delta,m}(t',v)\textrm{ as }l\to +\infty,\label{st7t}
\end{equation}
for $t'\in \R$ and $v\in V$ such that $v$ and $v_0'$ are linearly independent, where
\begin{eqnarray}
f_{\delta,m}(t',v)&:=&
{1\over |v_0'|}(k-\tilde k)(x_0'+t'\hat{v_0'},v_0',v)\rho_m(t',v)\label{st7tb}\\
&&\times\left[\chi_{{\mathcal V}_\delta}(x+R\hat v,v)E(x,t,v,v_0')\right]_{t=t(x_0',v_0',t',v)
\atop x=x(x_0',v_0',t',v)},\nonumber
\end{eqnarray}
for $t'\in \R$ and $v\in V$ such that $v$ and $v_0'$ are linearly
independent.  Therefore we obtain by the Lebesgue dominated
convergence theorem that
\begin{equation}
\lim_{l\to +\infty}I_2^1(\phi_{\delta,m,l})=\int_V\int_{-R}^Rf_{\delta,m}(t',v)dt'dv.\label{st7u}
\end{equation}

Let $0<\delta<\min(R,\v_0^{-1})$.  We also have $f_{\delta,m}(t',v)\le
\v_0^{-1}(k+\tilde k)(x_0'+t'\hat{v_0'},v_0',v))$, for $m\in \N$.
From \eqref{st7u}, \eqref{st7tb} and \eqref{st7u1}, it follows that
\begin{eqnarray}
&&\lim_{m\to+\infty}\lim_{l\to +\infty}I_2^1(\phi_{\delta,m,l})=\int_V\int_{-R}^R
{1\over |v_0'|}|k-\tilde k|(x_0'+t'\hat{v_0'},v_0',v)\label{st7u2}\\
&&\left[\chi_{{\mathcal V}_\delta}(x+R\hat v,v)E(x,t,v,v_0')
\right]_{t=-(x_0'+t_0'\hat{v_0'})\hat v\atop x-t\hat v=x_0'+t'\hat{v_0'}}dt'dv.\nonumber
\end{eqnarray}
From \eqref{st7u2}, we deduce
\begin{eqnarray}
\lim_{\delta\to 0^+}\lim_{m\to+\infty}\lim_{l\to +\infty}I_2^1(\phi_{\delta,m,l})&=&\!\!\!\!\int_V\int_{-R}^R
{1\over |v_0'|}|k-\tilde k|(x_0'+t'\hat{v_0'},v_0',v)\label{st7u3}\\
&&\qquad \times E(x,v,t,v_0')\!\!\!_{t=-(x_0'+t'\hat{v_0'})\hat{v}\atop x=x_0'+t'\hat{v_0'}-((x_0'+t'\hat{v_0'})\hat{v})\hat{v}}\!\!\!\!\!\!\!\!\!dt'dv.\nonumber
\end{eqnarray}
From \eqref{l5.2d}, it follows that
\begin{equation}
|I_2^2(\phi_{\delta,m,l})|\le {2R\over \v_0}\|\tilde \sigma_p(x_0'+t'\hat{v_0'},{v_0'})\|_{L^\infty(\R_{t'})}\sup_{(x,v)\in F\atop t\in \R}\left|(E-\tilde E)(x,t,v,v_0')\right|,
\label{st8a}
\end{equation}
for $0<\delta<\min(R,\v_0^{-1})$, $m\in \N$ and $l\in \N$,
$l>(R-\delta)^{-1}+\delta$.  Using \eqref{st1} and \eqref{st2d}, we
obtain
\begin{equation}
|I_2(\phi_{\delta,m,l},\ep)|\le\|\A-\tilde\A\|_{{\mathcal L}(L^1(F_-),L^1(F_+))}+|I_1(\phi_{\delta,m,l},\ep)|+|I_3(\phi_{\delta,m,l},\ep)|,\label{st8b}
\end{equation}
for  $0<\delta<\min(R,\v_0^{-1})$, $m\in \N$ and $l\in \N$, $l>(R-\delta)^{-1}+\delta$.
From \eqref{st8b}, \eqref{l5.2a} and \eqref{l5.2b}, it follows that
\begin{eqnarray}
|I_2^1(\phi_{\delta,m,l})|&\le&\|\A-\tilde\A\|_{{\mathcal L}(L^1(F_-),L^1(F_+))}
+|I_2^2(\phi_{\delta,m,l})|+|\lim_{\ep\to 0^+}I_1(\phi_{\delta,m,l},\ep)|\nonumber\\
&&+\limsup_{\ep\to 0^+}|I_3(\phi_{\delta,m,l},\ep)|,\label{st8c}
\end{eqnarray}
for $0<\delta<\min(R,\v_0^{-1})$, $m\in \N$ and $l\in \N$,
$l>(R-\delta)^{-1}+\delta$.  Estimate \eqref{t5.1b} follows from
\eqref{st8c}, \eqref{st7j}, \eqref{st7q}, \eqref{l5.2b}, \eqref{st7u3}
and \eqref{st8a}.
\end{proof}

\begin{proof}[Proof of Theorem \ref{thm:stab2}.]
  The method we use to prove \eqref{t5.2a} is the same as in
  \cite{W-AIHP-99}.  Let $(\sigma,k)$, $(\tilde \sigma,\tilde k)\in
  {\mathcal M}$.  Let $f=\sigma-\tilde\sigma$ and consider $Pf$ the X-ray
  transform of $f=\sigma-\tilde\sigma$ defined by
  $Pf(x,\theta):=\int_{-\infty}^{+\infty}f(t\theta+x)dt$ for a.e.
  $(x,\theta)\in T\S^{n-1}$.

From \eqref{eq:hyp2} and $f_{|X}\in H^{{n\over 2}+\tilde r}(X)$, it follows that 
\begin{equation}
\|f\|_{H^{-{1\over 2}}(X)}\le D_1(n,X)\|Pf\|_{*},\label{t3.2p1}
\end{equation}
where 
$$
\|Pf\|_{*}:=\left(\int_{\S^{n-1}}\int_{\Pi_\theta}|Pf(x,\theta)|^2dx d\theta\right)^{1\over 2}
$$
and $D_1(n,X)$ is a real constant which does not depend on $f$ and  $\Pi_\theta:=\{x\in \R^n\ |\ x\theta=0\}$  for $\theta\in \S^{n-1}$.
Using \eqref{eq:hyp2} (and $(\sigma,k)$, $(\tilde \sigma, \tilde k)\in {\mathcal M}$),  it follows that $Pf(x,\theta)=0$ for $(x,\theta)\in T\S^{n-1}$ and $|x|\ge R$. Therefore using also \eqref{t3.2p1} we obtain 
\begin{equation}
\|f\|_{H^{-{1\over 2}}(X)}\le D_2(n,X)\|Pf\|_{L^\infty(T\S^{n-1})},\label{t3.2p2}
\end{equation}
where $D_2(n,X)$ is a real constant which does not depend on $\sigma$.

We also use the following interpolation inequality:
\begin{equation}
\|f\|_{H^s(X)}\le \|f\|_{H^{{n\over 2}+\tilde r}}^{2s+1\over n+1+2\tilde r}\|f\|_{H^{-{1\over 2}}}^{n+2\tilde r\over n+1+2\tilde r},\label{t3.2p2b}
\end{equation}
for $-{1\over 2}\le s\le {n\over 2}+\tilde r$.  As $\sigma\in {\mathcal
  M}$, it follows that
\begin{equation}
\|\sigma\|_\infty\le D_3(n,\tilde r)\|\sigma\|_{H^{{n\over 2}+\tilde r}}\le D_3(n,\tilde r)M.\label{t3.2p3}
\end{equation}
Therefore,
\begin{equation}
\int_{-R}^R\sigma (x_0'-s\hat {v_0'})ds\le 2RD_3(n,\tilde r)M,\label{t3.2p4}
\end{equation}
for a.e. $(x_0',\hat{v_0'})\in T\S^{n-1}$.  From \eqref{t3.2p4} it
follows that
\begin{eqnarray}
&&\left|e^{-|v_0'|^{-1}\int_{-R}^R\sigma (x_0'-s\hat {v_0'},v_0')ds }
-e^{-|v_0'|^{-1}\int_{-R}^R\tilde\sigma (x_0'-s\hat {v_0'},v_0')ds }\right|\nonumber\\
&&\ge {e^{-2\v_0^{-1}RD_3(n,\tilde r)M}\over \V_0}|P(\sigma-\tilde \sigma)(x_0',\hat{v_0'})|,\label{t3.2p5}
\end{eqnarray}
for a.e. $(x_0', v_0')\in \R^n\times V$, $x_0'v_0'=0$ (we used the
equality $e^{t_1}-e^{t_2}=e^{c}(t_2-t_1)$ for $t_1<t_2\in \R$ and for
some $c\in [t_1,t_2]$ which depends on $t_1$ and $t_2$).  (In fact,
the estimate \eqref{t3.2p5} is valid for any $(x_0', v_0')\in
\R^n\times V$, $x_0'v_0'=0$, such that $\{x_0'+tv_0'\ |\ t\in \R\}\cap
X\not=\emptyset$ or $\{x_0'+tv_0'\ |\ t\in \R\}\cap \bar
X=\emptyset$.)  Combining \eqref{t3.2p5}, \eqref{t3.2p2}, and
\eqref{t5.1a}, we obtain
\begin{equation}
{e^{-2\v_0^{-1}RD_3(n,\tilde r)M}\over D_2(n,X)\V_0}\|\sigma-\tilde \sigma\|_{H^{-{1\over 2}}(X)}\le \|\A-\tilde\A\|_{{\mathcal L}(L^1(F_-),L^1(F_+))}.\label{t3.2p6}
\end{equation}
Combining \eqref{t3.2p6} and \eqref{t3.2p2b}, we obtain \eqref{t5.2a}.

We now prove \eqref{t5.2b}.  Using $|v_0'|^{-1}\ge \V_0^{-1}$ for
$v\in V$, \eqref{st2c0}, and \eqref{t3.2p3}, we obtain that
$$
{1\over
  |v_0|'}\int_V\int_{-R}^R\left|k(x_0'+t'\hat{v_0'},v_0',v)-\tilde
  k(x_0'+t'\hat{v_0'},v_0',v)\right|
{E(x,v,t,v_0')}_{t=t(x_0',v_0',t',v)\atop x=x(x_0',v_0',t',v)}dt'dv
$$
\begin{equation}
\ge {e^{-4\v_0^{-1}RD_3(n,\tilde r)M}\over \V_0}\int_V\int_{-R}^R|k-\tilde k|(x_0'+t'\hat{v_0'},v_0',v)|dt'dv,\label{t3.2p7}
\end{equation}
for any $(x_0',v_0')\in \R^n\times V$, $x_0'v_0'=0$, such that
$x_0'+sv_0'\in X$ for some $s\in \R$, and where
$(t(x_0',v_0',t',v),x(x_0',v_0',t',v))$ is defined by \eqref{l5.2da}
for $v\in V$ and $t'\in (-R,R)$.

As $(\tilde \sigma,\tilde k)\in {\mathcal M}$ we have $\|\tilde
\sigma_p\|_{\infty}\le M$. Using the latter estimate, \eqref{st2c0},
and $|v|^{-1}\le\v_0^{-1}$ for all $v\in V$, we obtain
\begin{eqnarray}
&&\|\tilde \sigma_p(x_0'+t'\hat{v_0'},{v_0'})\|_{L^\infty(\R_{t'})}\sup\limits_{(x,v)\in F\atop t\in \R}|E-\tilde E|(x,v,t,v_0')\le Me^{4\v_0^{-1}RD_3(n,\tilde r)M}\nonumber\\
&&\times\sup\limits_{(x,v)\in F\atop t\in \R}\left[{1\over |v|}\int_{-R}^t\!\!\!\!\!|\sigma-\tilde \sigma|(x-s\hat v,v)ds
+{1\over |v_0'|}\int_0^{R+(x-t\hat v)\hat{v_0'}}\!\!\!\!\!\!\!\!\!\!\!\!\!\!\!\!\!\!\!|\sigma-\tilde \sigma|(x-t\hat v-s\hat{v_0'},{v_0'})ds\right]\nonumber\\
&&\le 4R\v_0^{-1}Me^{4\v_0^{-1}RD_3(n,\tilde r)M}\|\sigma-\tilde \sigma\|_{\infty},\label{t3.2p8}
\end{eqnarray}
for any $(x_0',v_0')\in \R^n\times V$, $x_0'v_0'=0$, such that
$x_0'+t'v_0'\in X$ for some $t'\in \R$. (We also used $|e^u-e^{\tilde
  u}|\le e^{\max(|u|, |\tilde u|)}|u-\tilde u|$ where
$u=-|v|^{-1}\int_{-R}^t\sigma(x-s\hat v,v)
ds-|v_0'|^{-1}\int_0^{R+(x-t\hat v)\hat{v_0'}}\tilde\sigma(x-t\hat
v-s\hat{v_0'},{v_0'})ds$ and $\tilde u$ denotes the real number
obtained by replacing $\sigma$ by $\tilde \sigma$ on the right-hand
side of the latter equality which defines $u$; using \eqref{t3.2p3}
(for $\sigma$ and for $\tilde \sigma$) we obtain $\max(|u|,|\tilde
u|)\le 4R\v_0^{-1}D_3(n,\tilde r)M$.)  Note that
$\|\sigma-\tilde\sigma\|_\infty\le D_3(n,r)\|\sigma-\tilde
\sigma\|_{H^{{n\over 2}+r}}$ for $0<r<\tilde r$ (see \eqref{t3.2p3}).
Therefore, combining \eqref{t3.2p7}, \eqref{t3.2p8}, \eqref{t5.1b} and
\eqref{t5.2a}, we obtain \eqref{t5.2b}.

Let us finally prove \eqref{t5.2c}. Let $0<r<\tilde r$ and let
$\theta={2(\tilde r-r)\over n+1+2 \tilde r}$. From \eqref{t5.2b} it
follows that
\begin{eqnarray}
&&\int_{x_0'v_0'=0\atop |x_0'|<R}\int_{-R}^R\int_V\left|k(x_0'+t'\hat{v_0'},v_0',v)-\tilde k(x_0'+t'\hat{v_0'},v_0',v)\right|dvdt'dx_0'\nonumber\\
&\le&
D_4\|\A-\tilde\A\|_{{\mathcal L}(L^1(F_-),L^1(F_+))}^\theta
\left(1+\|\A-\tilde\A\|_{{\mathcal L}(L^1(F_-),L^1(F_+))}^{1-\theta}\right),\label{t3.2p9}
\end{eqnarray}
where $D_4=C_2\int_{x_0'v_0'=0\atop |x_0'|<R}dx_0'$ and $C_2$ is the
constant that appears on the right-hand side of \eqref{t5.2b}.  From
\eqref{eq:hyp2}, \eqref{t3.2p9} and the change of variables
``$x=x_0'+t'v_0'$'', it follows that
\begin{eqnarray}
&&\int_{\R^n\times V}\left|k(x,v_0',v)-\tilde k(x,v_0',v)\right|dvdx
\nonumber\\
&\le&
D_4\|\A-\tilde\A\|_{{\mathcal L}(L^1(F_-),L^1(F_+))}^\theta
\left(1+\|\A-\tilde\A\|_{{\mathcal L}(L^1(F_-),L^1(F_+))}^{1-\theta}\right).\label{t3.2p10}
\end{eqnarray}
Integrating on $v_0'\in V$ both sides of \eqref{t3.2p10}, we obtain \eqref{t5.2c}.

\end{proof}

%%%%%%%%%%%%%%%%%%%%%%%%%%%%%%%%%%%%%%%%%%%%%%%%%%%%%%%%%%%%%%%%%%%%%%%%%%%%%%%%%%%%%%%%%%%%

\section{Decomposition of the albedo operator}
\label{sec:proof2}
We now prove Lemmas \ref{lem:dec1} and \ref{lem:dec2}.
\begin{proof}[Proof of Lemma \ref{lem:dec1}.]
  Let $\psi_-\in L^1(O, |v|dx dv)$. Using the definition of $K$, we
  obtain
\begin{eqnarray}
({\mathcal R}\psi_-)(x,v)&=&\left(K^2\psi_-\right)_{|F_+}(x,v)\label{SP6}\\
&=&\int_{V\times V}{1\over |v|}{1\over |v_1|}\int_0^{2R}\int_0^{R+(x-t\hat v)\hat v_1}k(x-t\hat v,v_1,v)\nonumber\\
&&\times k(x-t\hat v-t_1\hat{v_1},v',v_1)E_0(x,v,x-t\hat v,v_1,x-t\hat v-t_1\hat v_1)\nonumber\\
&&\times \psi_-(x-t\hat v-t_1\hat{v_1},v')dt_1dtdv'dv_1,\nonumber
\end{eqnarray}
for a.e. $(x,v)\in F_+$, where
\begin{equation}
E_0(x,v,x-t\hat v,v_1,x-t\hat v-t_1\hat v_1)=e^{-|v|^{-1}\int_0^t\sigma(x-s\hat v,v)ds-|v_1|^{-1}\int_0^{t_1}\sigma (x-t\hat v-s\hat{v_1},v_1)ds },\label{SP6aa}
\end{equation}
for $x\in \R^n$, $t$, $t_1\in \R$ and $v$, $v_1\in V$. We recall that ${\mathcal R}$ is a bounded
operator from $L^1(O, |v|dx dv)$ to $L^1(F_+)$, i.e.
\begin{equation}
\|{\mathcal R}\psi\|_{F_+}\le C\||v|\psi\|_O, \textrm{ for any }\psi\in L^1(O, |v|dx dv).\label{SP6a}
\end{equation} 
Hence we obtain, in particular, that the integral in $t$, $t_1$, $v'$ and $v_1$, on the right-hand side of 
\eqref{SP6} is absolutely convergent for a.e. $(x,v)\in F_+$.

Let us assume first that $V=\S^{n-1}$. Performing the changes of
variables ``$x'=x-tv-t_1v_1$'' (``$dx'=t_1^{n-1}dt_1dv_1$''), we
obtain
\begin{equation}
({\mathcal R}\psi_-)(x,v)=\int_O\beta(x,v,x',v')\psi_-(x',v')dx'dv',\label{SP7a}
\end{equation}
where
\begin{equation}
  \label{SP7b}
  \beta(x,v,x',v') :=\phantom{wwwwwwwwwwwwwwwwwwwwwwwwwwwwwwwwwwwwwwww}
\end{equation}
\vspace{-.5cm}
%$$
%\beta(x,v,x',v')\phantom{wwwwwwwwwwwwwwwwwwwwwwwwwwwwwwwwwwwwwwww}
%$$
$$
\int_0^{2R}\left[{k(x-tv,v_1,v)k(x',v',v_1)\over |x-tv -x'|^{n-1}}
E_0(x,v,x-t v, v_1, x-t v-t_1 v_1)\right]_{t_1=|x-tv-x'|\atop v_1={x-t v-x'\over t_1}}\!\!\!\!\!\!dt,%\label{SP7b}
$$
for a.e. $(x,v)\in F_+$, $(x',v')\in O$, where $E_0$ is defined by \eqref{SP6aa}.

Now assume that $V$ is  an open subset of  $\R^n$, which satisfies 
$\v_0=\inf_{v\in V}|v|>0$.  
From \eqref{SP6}, it follows that
\begin{eqnarray}
({\mathcal R}\psi_-)(x,v)&=&\left(K^2\psi_-\right)_{|F_+}(x,v)\nonumber\\
&=&\int_{V\times \S^{n-1}}\!\!\!\!\!\!|v|^{-1}\int_{\v_0}^{+\infty}r^{n-2}\chi_V(r\omega)\int_0^{2R}\int_0^{R+(x-t\hat v)\omega}\!\!\!\!\!\!\!\!\!k(x-t\hat v,r\omega,v)\nonumber\\
&&\times k(x-t\hat v-t_1\omega,v',r\omega)E_0(x,v,x-t\hat v,r\omega,x-t\hat v-t_1\omega)\nonumber\\
&&\times \psi_-(x-t\hat v-t_1\omega,v')dt_1dtdr dv'd\omega,\label{SP7be}
\end{eqnarray}
for $(x,v)\in F_+$, where for $r>0$ and $\omega \in \S^{n-1}$.
Performing the changes of variables  ``$x'=x-t\hat v-t_1\omega$'' (``$dx'=t_1^{n-1}dt_1d\omega$''), we obtain
\begin{equation}
({\mathcal R}\psi_-)(x,v)=\int_O\beta(x,v,x',v')\psi_-(x',v')dx'dv',\label{SP7c}
\end{equation}
where
\begin{eqnarray}
\beta(x,v,x',v')&=&{1\over|v|}\int_{\v_0}^{+\infty}r^{n-2}\int_0^{2R}\left[\chi_V(r\omega){k(x-t\hat v,r\omega,v)k(x',v',r\omega)\over |x-x'-t\hat v|^{n-1}}\right.\nonumber\\
&&\left.\times E_0(x,v,x-t\hat v,r\omega,x-t\hat v-t_1\omega)\right]_{t_1=|x-x'-t\hat v|\atop t_1\omega=x-x'-t\hat v}dtdr,\label{SP7d}
\end{eqnarray}
for a.e. $(x,v)\in F_+$, $(x',v')\in O$, where $E_0$ is defined by \eqref{SP6aa}.

From \eqref{SP6a}, \eqref{SP7a}--\eqref{SP7b}, and
\eqref{SP7c}--\eqref{SP7d}, it follows that for a.e. $(x',v')\in O$,
$\beta(x,v,$ $x',v')\in L^1(F_+)$.  Moreover from \eqref{SP6a}, it
follows that the function $O\ni(x',v')\to
\beta(x,v,x',v')\psi(x',v')\in L^1(F_+)$ belongs to $L^1(O,|v|dx dv)$
for any $\psi\in L^1(O, |v'|dx' dv')$.  Therefore
\begin{equation}
|v'|^{-1}\beta\in L^{\infty}(O, L^1(F_+)).\label{SP9}
\end{equation}

Now we prove \eqref{S0cb}.  Assume $k\in L^\infty(\R^n\times V\times
V)$ and let $1<p<1+{1\over n-1},$ ${p'}^{-1}+p^{-1}=1,$ be fixed for
the rest of the proof of Lemma \ref{lem:dec1}.  We use \eqref{SP23d}.
Using H\"older estimate, the change of variables ``$y=x-t\hat v$''
($dy=dxdt$) and the spherical coordinates, we obtain
$$
\int_{V_\delta}\int_{R\hat v+\Pi_v(R)}\left(\int_0^{2R}{1\over |x-x'-t\hat v|^{n-1}}dt\right)^p 
dxdv\phantom{wwwwwwwwwwwwwwwww}
$$
$$
\le (2R)^{p\over p'}\int_{V_\delta}\int_{R\hat v+\Pi_v(R)}\int_0^{2R}{dt 
dxdv\over |x-x'-t\hat v|^{p(n-1)}}\phantom{wwwwwwwwwwwwwwwww}
$$
\begin{equation}
=(2R)^{p\over p'}\int_{V_\delta}\int_{y\in \R^n\atop (y,v)\in O}{dydv\over |y-x'|^{p(n-1)}}
\le (2R)^{p\over p'}{\rm Vol}(V_\delta){\rm Vol}(\S^{n-1}){(4R)^{n-(n-1)p}\over n-(n-1)p},\label{SP23d}
\end{equation}

for $x'\in \R^n,$ $|x'|<2R$, where
\begin{equation}
V_\delta:=\{v\in V\ |\ |v|<\delta^{-1}\}.\label{SP23e}
\end{equation}

Assume first that $V=\S^{n-1}$ and let $\phi$ be a continuous function on $F_+$.
Then using \eqref{SP7b}, $\sigma\ge 0$, H\"older estimate and \eqref{SP23d} (with $\delta={1\over 2}$), we obtain
\begin{eqnarray}
&&\!\!\!\!\!\!\!\!\!\!\left|\int_V\int_{R\hat v+\Pi_v(R)}\!\!\!\!\!\!\!\!\phi(x,v)\beta(x,v,x',v')dxdv\right|\le\|k\|_\infty^2
\int_V\int_{R\hat v+\Pi_v(R)}\!\!\!\!\!\!\!\!\!\!\!\!\!\!\!\!\!\!|\phi(x,v)|\int_0^{2R}\!\!\!\!{dt dx dv\over |x-tv -x'|^{n-1}}\nonumber\\
&&\!\!\!\!\!\!\!\!\!\!\le\|k\|_\infty^2\left(\int_V\int_{R\hat v+\Pi_v(R)}\!\!\!\!\!\!\!\!\!\!\!\!\!\!\!\!\!\!|\phi(x,v)|^{p'}dx dv\right)^{1\over p'}\nonumber\\
&&\times\left(\int_V\int_{R\hat v+\Pi_v(R)}\left(\int_0^{2R}{1\over |x-t v -x'|^{n-1}}dt\right)^p dx dv
\right)^{1\over p}\nonumber\\
&&\!\!\!\!\!\!\!\!\!\!\le C\left(\int_V\int_{R\hat v+\Pi_v(R)}\!\!\!\!\!\!\!\!\!\!\!\!\!\!\!\!\!\!|\phi(x,v)|^{p'}dx dv\right)^{1\over p'}, \label{SPn1}
\end{eqnarray}
where $C=(2R)^{1\over p'}\|k\|_\infty^2{\rm Vol}(\S^{n-1})^{2\over p}\left({(4R)^{n-(n-1)p}\over n-(n-1)p}\right)^{1\over p}$, which proves \eqref{S0cb} for $V=\S^{n-1}$.

Now assume that $V$ is an open subset of $\R^n$ which satisfies $\inf_{v\in V}|v|>0$.
Let $\ep'>0$ an $\delta >0$ be positive real numbers. 
Let $\phi$ be a compactly  supported and continuous function on $F_+$ such that
${\rm supp}\phi\subseteq\{(x,v)\in F_+\ |\ |v|<\delta^{-1}\}$.
We use the following lemma, whose proof is postponed to the 
end of this section.
\begin{lemma}
  \label{lem:beta1}
The nonnegative measurable function $\beta_1$ defined for a.e. $(x,v,x',v')$ $\in F_+\times O$ by
\begin{eqnarray}
\beta_1(x,v,x',v')&=&{1\over |v|}\int_{\v_0}^{+\infty}r^{n-1}\int_0^{2R}\left[\chi_V(r\omega){k(x-t\hat v,r\omega,v)k(x',v',r\omega)\over |x-x'-t\hat v|^{n-1}}\right.\nonumber\\
&&\left.\times E_0(x,v,x-t\hat v,r\omega, x-t\hat v -t_1\omega)\right]_{t_1=|x-x'-t\hat v|\atop t_1\omega=x-x'-t\hat v}dtdr,\label{SP20}
\end{eqnarray}
belongs to $L^\infty(O,L^1(F_+))$, where $E_0$ is defined by \eqref{SP6aa}.
\end{lemma}
%The proof of Lemma \ref{lem:beta1} is given below at the end of this
%Section.  
Let $M_{\ep'}>\v_0$ be defined by
\begin{equation}
M_{\ep'}=\v_0+{\ep'}^{-1}\left\|\int_V\int_{R\hat v+\Pi_v(R)}\beta_1(x,v,x',v')|v|dx dv\right\|_{L^\infty(O)}.\label{SP21}
\end{equation}
From \eqref{SP7d}, it follows that
\begin{equation}
\int_V\int_{R\hat v+\Pi_v(R)}\phi(x,v)|v|\beta(x,v,x',v')dxdv=I_1(x',v')+I_2(x',v'),\label{SP22}
\end{equation}
for a.e. $(x',v')\in O$ and where
\begin{eqnarray}
I_1(x',v')&=&\int_{V_\delta}\int_{R\hat v+\Pi_v(R)}\phi(x,v)\label{SP22b}\\
&&\int_{\v_0}^{M_{\ep'}}r^{n-2}\int_0^{2R}\left[\chi_V(r\omega){k(x-t\hat v,r\omega,v)k(x',v',r\omega)\over |x-x'-t\hat v|^{n-1}}\right.\nonumber\\
&&\left.\times E_0(x,v, x-t\hat v, r\omega,x-t\hat v-t_1\omega)\right]_{t_1\omega=x-x'-t\hat v}dt dr dx dv,\nonumber\\
I_2(x',v')&=&\int_{V_\delta}\int_{R\hat v+\Pi_v(R)}\phi(x,v)\label{SP22c}\\
&&\int_{M_{\ep'}}^{+\infty}r^{n-2}\int_0^{2R}\left[\chi_V(r\omega){k(x-t\hat v,r\omega,v)k(x',v',r\omega)\over |x-x'-t\hat v|^{n-1}}\right.\nonumber\\
&&\left.\times E_0(x,v, x-t\hat v, r\omega,x-t\hat v-t_1\omega)\right]_{t_1\omega=x-x'-t\hat v}dt dr dx dv.
\nonumber
\end{eqnarray}
Using \eqref{SP22c} and the estimates $r^{n-2}=r^{n-1}r^{-1}\le M_{\ep'}^{-1}r^{n-1}$ for $v\in V$ and $r\ge M_{\ep'}$, and using \eqref{SP21}, we obtain
\begin{eqnarray}
|I_2(x',v')|&\le&\|\phi\|_{L^\infty(F_+)}M_{\ep'}^{-1}\left\|\int_{v\in V}\int_{R\hat v+\Pi_v(R)}\beta_1(x,v,x'',v'')|v|dx dv\right\|_{L^\infty(O)}\nonumber\\
&\le&\ep'\|\phi\|_{L^\infty(F_+)}, \textrm{ for a.e. }(x',v')\in F_+.\label{SP23b}
\end{eqnarray}
From \eqref{SP22b} and H\"older estimate, it follows that
\begin{eqnarray}
|I_1(x',v')|&\!\!\!\le&\!\!\!\|k\|_{\infty}^2\int_{V_{\delta}}\int_{R\hat v+\Pi_v(R)}\!\!\!\!\!\!\!\!\!\!\!\!\!\!\!\!\!|\phi(x,v)|
\int_{\v_0}^{M_{\ep'}}r^{n-2}dr\int_0^{2R}{1\over |x-x'-t\hat v|^{n-1}}dt dx dv.\nonumber\\
&\!\!\!\le&\!\!\!\|k\|_{\infty}^2{M_{\ep'}^{n-1}\over n-1}\left(\int_{V_\delta}\int_{R\hat v+\Pi_v(R)}\left(\int_0^{2R}{1\over |x-x'-t\hat v|^{n-1}}dt\right)^p 
dxdv\right)^{1\over p}\nonumber\\
&&\!\!\!\times
\left(\int_{V_\delta}\int_{R\hat v+\Pi_v(R)}|\phi(x,v)|^{p'}dxdv\right)^{1\over p'},\label{SP23c}
\end{eqnarray}
for a.e. $(x',v')\in O$.  Combining \eqref{SP22},
\eqref{SP23b}--\eqref{SP23c} and \eqref{SP23d}, we obtain \eqref{S0cb}
with
$$
C(\ep',\delta,p)=\|k\|_{\infty}^2{M_{\ep'}^{n-1}\over n-1}(2R)^{1\over p'}\left({\rm Vol}(V_\delta){\rm Vol}(\S^{n-1})\right)^{1\over p}
\left({(4R)^{n-(n-1)p}\over n-(n-1)p}\right)^{1\over p}.
$$

\end{proof}

\begin{proof}[Proof of Lemma \ref{lem:dec2}.]
Let $\phi_-\in C^1_0(F_-)$ (which denotes the spaces of $C^1$ compactly supported  functions  on $F_-$). Let $\phi:=(I+K)^{-1}J\phi_-$. Then
note that
\begin{equation}
\phi:=J\phi_--KJ\phi_-+K^2(I+K)^{-1}J\phi_-.\label{SP1}
\end{equation}
Thus
\begin{equation}
\A\phi_-=\phi_{|F_+}:=\left(J\phi_-\right)_{|F_+}-\left(KJ\phi_-\right)_{|F_+}+{\mathcal R}(I+K)^{-1}J\phi_-.\label{SP3}
\end{equation}
From \eqref{2.3} and \eqref{S2a}, it follows that
\begin{equation}
\left(J\phi_-\right)_{|F_+}(x,v)=\int_V\int_{x'v'=0\atop |x'|<R}\alpha_1(x,v,x',v')\phi_-(x'-R\hat v',v')dx' dv',\ (x,v)\in F_+.\label{SP4}
\end{equation}
From the definitions of $K$ and $J$, we obtain
\begin{equation}
-\left(KJ\phi_-\right)_{|F_+}(x,v)=\int_V\int_{x'v'=0\atop |x'|<R}\alpha_2(x,v,x',v')\phi_-(x'-R\hat v',v')dx' dv',\ (x,v)\in F_+.\label{SP5}
\end{equation}
Lemma \ref{lem:dec2} follows from \eqref{SP3}--\eqref{SP5} and Lemma \ref{lem:dec1}.
\end{proof} 

\begin{proof}[Proof of Lemma \ref{lem:beta1}]
  Using \eqref{SP20} and the estimate $\sigma\ge 0$ and using the
  change of variables ``$y=x-t\hat v$'' ($dy=dtdx$) and
  \eqref{eq:hyp1}, and spherical coordinates, we obtain
$$
\int\limits_{v\in V}\int\limits_{R\hat v+\Pi_v(R)}|v|\beta_1(x,v,x',v')dxdv\phantom{wwwwwwwwwwwwwwwwwwwwwwwwwww}
$$
$$
\le\int\limits_V\int\limits_{R\hat v+\Pi_v(R)}
\int\limits_{\v_0}^{+\infty}r^{n-1}\int\limits_0^{2R}\left[\chi_V(r\omega){k(x-t\hat v,r\omega,v)k(x',v',r\omega)\over |x-x'-t\hat v|^{n-1}}\right]_{\omega={x-t\hat v-x'\over |x-t\hat v-x'|}}
\!\!\!\!\!\!\!\!\!\!\!\!\!\!\!\!\!dtdrdxdv,
$$
$$
=\int\limits_V\int\limits_{y\in \R^n\atop |y|<R}{1\over |x'-y|^{n-1}}
\int\limits_{\v_0}^{+\infty}r^{n-1}\left[\chi_V(r\omega)k(y,r\omega,v)k(x',v',r\omega)\right]_{\omega={y-x'\over |y-x'|}}drdydv
$$
$$
=\int\limits_{\S^{n-1}}\int_0^R\chi_{|y|<R}(x'+r'\omega)
\int\limits_{\v_0}^{+\infty}r^{n-1}\chi_V(r\omega)\sigma_p(x'+r'\omega,r\omega)k(x',v',r\omega)drdr'd\omega
$$
$$
\le\|\sigma_p\|_\infty R\int\limits_{\S^{n-1}}
\int\limits_{\v_0}^{+\infty}r^{n-1}\chi_V(r\omega)k(x',v',r\omega)drd\omega =R\|\sigma_p\|_\infty \sigma_p(x',v')\le R\|\sigma_p\|_\infty^2,
$$
for a.e. $(x',v')\in O$.  The lemma is proved.
\end{proof}

%%%%%%%%%%%%%%%%%%%%%%%%%%%%%%%%%%%%%%%%%%%%%%%%%%%%%%%%%%%%%%%%%%%%%%%%%%%

\section{Proof of existence of the albedo operator}
\label{sec:proof1}
In this section, we prove Lemma \ref{lem:exist} and Propositions
\ref{prop:fwd} and \ref{prop:albedo}.
\begin{proof}[Proof of Lemma \ref{lem:exist}]
  Using the definition of $\Tu^{-1}$, the estimate $\sigma\ge 0$ and
  \eqref{2.1}, we have
\begin{eqnarray*}
\||v|\Tu^{-1}f\|_O&\le&\int_V\int_{\Pi_v(R)}\int_{-R}^R\int_0^{R+w}|f(y+(w-t)\hat v,v)|dtdwdydv\\
&=&\int_V\int_{\Pi_v(R)}\int_{-R}^R\int_{-R}^w|f(y+t\hat v,v)|dtdwdydv\le 2R\|f\|_O,
\end{eqnarray*}
for $f\in L^1(O)$.

Using the  definition of $A_2$ and \eqref{2.1}, we have
\begin{eqnarray*}
\|A_2|v|^{-1}f\|_O&\le&\int_{V\times V}\int_{\Pi_v(R)}\int_{-R}^Rk(y+t\hat v,v',v)|v'|^{-1}|f(y+t\hat v,v')|dtdydvdv'\\
&=&\int_V\int_{\Pi_{v'}(R)}\int_{-R}^R\sigma_p(y'+t'\hat v',v')|v'|^{-1}|f(y'+t'\hat v',v')| dt'dy'dv'\\
&&\le \||v'|^{-1}\sigma_p(x',v')\|_{L^\infty(O)}\|f\|_O,
\end{eqnarray*}
for $f\in L^1(O)$. We also used \eqref{eq:hyp1} and the change of
variables
\begin{equation}
\int_{yv=0}\int_{-\infty}^{+\infty}f(y+t\hat v)dtdy=\int_{y'v'=0}\int_{-\infty}^{+\infty}f(y'+t'\hat v')dt'dy',\label{P2}
\end{equation}
for $f\in L^1(\R^n)$ and $v,v'\in V$.

Using the definition of $\Tu^{-1}$ and $A_2$ and Lemma
\ref{lem:chvar}, \eqref{c2} and \eqref{P2}, we obtain
$$
\|A_2\Tu^{-1} f\|_O\phantom{wwwwwwwwwwwwwwwwwwwwwwwwwwwwwwwwwwwwwwww}
$$
\begin{eqnarray}
\le\int_{V\times V}{1\over |v'|}\int_{\Pi_v(R)}\int_{-R}^Rk(y+w\hat v,v',v)\int_0^{R+(y+w\hat v)\hat{v'}}
\!\!\!\!\!\!\!e^{-|v'|^{-1}\int_0^t\sigma_p(y+w\hat v-s\hat{v'},v')ds}\nonumber\\
\times|f(y+w\hat v-t\hat{v'},v')|dtdwdydvdv'
\nonumber
\end{eqnarray}
\begin{eqnarray}
=\int_{V\times V}{1\over |v'|}\int_{\Pi_{v'}(R)}\int_{-R}^Rk(y'+w'\hat{v'},v',v)\int_0^{R+w'}e^{-|v'|^{-1}\int_0^t\sigma_p(y'+(w'-s)\hat{v'},v')ds}\nonumber\\
\times|f(y'+(w'-t)\hat{v'},v')|dtdw'dy'dvdv'
\nonumber
\end{eqnarray}
$$
=\int_V\int_{\Pi_{v'}(R)}\int_{-R}^R\left(\int_t^R(-{d\over dw'}e^{-|v'|^{-1}\int_t^{w'}\sigma_p(y'+s\hat{v'},v')ds}dw'\right)|f(y'+t\hat{v'},v')|dtdy'dv'
$$
$$
=\int_O\left(1-e^{-|v'|^{-1}\int_0^{R-x\hat{v'}}\sigma_p(x+sv',v')ds}\right)|f(x',v')|dx'dv'\le(1-e^{-2R\v_0^{-1}\|\sigma_p\|_{\infty}})\|f\|_O,
$$
for $f\in L^1(O)$.

Item iii follows from items i and ii (under \eqref{c1}, we also use that $\|A_2\Tu^{-1}\|\le \|A_2|v|^{-1}\|\||v|\Tu^{-1}\|$). 
\end{proof}

\begin{proof}[Proof of Proposition \ref{prop:fwd}]
We first prove item i. 

Assume \eqref{c5}. For all $f\in D(\T)$,
\begin{equation}
\T f=(\Tu+A_2)f=(I+A_2\Tu^{-1})\Tu f.\label{P3.8}
\end{equation}
From \eqref{c5}  it follows that $\T$ admits a bounded inverse in $L^1(O)$ given by 
$\T^{-1}:=\Tu^{-1}(I+A_2\Tu^{-1})^{-1}$.
Using  the latter equality, we obtain 
\begin{eqnarray}
(I+K)(I-\T^{-1}A_2)
&\!\!\!\!\!=&\!\!\!\!\!I+\Tu^{-1}A_2-\Tu^{-1}(I+A_2\Tu^{-1})^{-1}A_2\nonumber\\
&&\!\!\!\!\!-\Tu^{-1}(I+A_2\Tu^{-1}-I)(I+A_2\Tu^{-1})^{-1}A_2=I.\label{P3.9}
\end{eqnarray}
The proof that $(I-\T^{-1}A_2)(I+K)=I$ is similar.  We now prove that
\eqref{c4} implies \eqref{c5}.  For $f\in D(\T)$,
\begin{equation}
\T f=(\Tu+A_2)f=\Tu(I+\Tu^{-1}A_2) f=\Tu(I+K)f.\label{P3.8iv}
\end{equation}
Let us prove $(I+K)(D(\T))=D(\T)$. From the latter equality and
\eqref{P3.8iv} it follows that $\T$ admits a bounded inverse in
$L^1(O)$ given by
\begin{equation}
\T^{-1}=(I+K)^{-1}\Tu^{-1}.\label{P3.9iv}
\end{equation} 
As $K=\Tu^{-1}A_2$, we have $(I+K)(D(\T))\subseteq D(\T)$.  Let $g\in
D(\T)$, and let $f=(I+K)^{-1}\Tu^{-1}g\in L^1(O)$. Then
$f=-K\Tu^{-1}g+g=-\Tu^{-1}A_2\Tu^{-1}g+g\in D(\T)$ (we recall that
$g\in D(\T)$).

Equality \eqref{P3.8} still holds. Using \eqref{P3.8}, \eqref{P3.9iv}
and the fact that $\Tu:D(\T)\to L^1(O)$ is one-to-one and onto
$L^1(O)$, we obtain \eqref{c5}.  Item i is thus proved.  Item ii
follows from item iii of Lemma \ref{lem:exist} and item i.  We shall
prove item iii.  Note that (see \eqref{P3.9})
\begin{equation}
(I+K)(I-\T^{-1}A_2)=I=(I-\T^{-1}A_2)(I+K) \textrm{ in }{\mathcal L}(L^1(O)).\label{P4}
\end{equation}
Note also that $L^1(O,|v|dx dv)\subseteq L^1(O)$ and recall that $K$
is a bounded operator in $L^1(O, |v|dx dv)$. Therefore, we only have to
prove that $\T^{-1}A_2$ defines a bounded operator in $L^1(O,|v|dx
dv)$.  Note that
\begin{equation}
\T^{-1}=\Tu^{-1}(I+A_2\Tu^{-1})^{-1}.\label{P5}
\end{equation}
From  item i, \eqref{P5} and item i of Lemma \ref{lem:exist}, it follows that $\T^{-1}A_2$ defines a bounded operator in $L^1(O,|v|dx dv)$. Thus
item iii is proved. 
\end{proof}

\begin{proof}[Proof of Proposition \ref{prop:albedo}.]
  Let $f_-\in L^1(F_-)$. From \eqref{2.4}, it follows that $Jf_-\in
  {\mathcal W}$. Hence $Jf_-\in L^1(O, |v|dx dv)$ and from \eqref{c3} it
  follows that \eqref{3.1} is uniquely solvable in $L^1(O, |v|dx dv)$
  and its solution is given by $(I+K)^{-1}Jf_-$ which satisfies
\begin{equation}
\|(I+K)^{-1}Jf_-\|_{L^1(O,|v|dx dv)}\le C_0\|f_-\|_{F_-},\label{P3.3}
\end{equation}
where $C_0=2R(1+\v_0^{-1}\|\sigma\|_{\infty})\|(I+K)^{-1}\|_{{\mathcal L}(L^1(O,|v|dx dv))}$.

Let $f:=(I+K)^{-1}Jf_-$. Hence by definition
\begin{equation}
f=Jf_--Kf \textrm{ in }L^1(O,|v|dx dv).\label{P3.4}
\end{equation}
Using \eqref{P3.4}, we check that the following equality is valid in
the sense of distributions:
\begin{equation}
T_0f=-A_1f-A_2f.\label{P3.5}
\end{equation}
Using \eqref{P3.5}, we obtain
$T_0f\in L^1(O)$ and 
\begin{equation}
\|T_0f\|\le (\|\sigma|v|^{-1}\|_{\infty}+\||v|^{-1}\sigma_p\|_{\infty})\|f\|_{L^1(O,|v|dx dv)}.\label{P3.6}
\end{equation}
Therefore $f\in {\mathcal W}$ (item i is thus proved), and using \eqref{2.2} and \eqref{P3.6} we obtain  
\begin{equation}
\|f_{|F_+}\|_{F_+}\le  \max((2R)^{-1},1)(\||v|^{-1}\sigma\|_{\infty}+\||v|^{-1}\sigma_p\|_{\infty}+1)\|f\|_{L^1(O,|v|^{-1}dx dv)}.\label{P3.7}
\end{equation}
Item ii follow from \eqref{P3.3} and \eqref{P3.7}.

\end{proof}

%%%%%%%%%%%%%%%%%%%%%%%%
\section*{Acknowledgments}
%%%%%%%%%%%%%%%%%%%%%%%%
This work was funded in part by the National Science Foundation under
Grants DMS-0239097 and DMS-0554097.

%%%%%%%%%%%%%%%%%
\bibliography{../../bibliography} 
\bibliographystyle{siam}

\end{document}